\documentclass[aps,prl,reprint,superscriptaddress]{revtex4-2}
\usepackage{graphicx}
\usepackage{amsfonts} 
\usepackage{dcolumn}
\usepackage{bm}
\usepackage{braket}
\usepackage{comment}
\usepackage{alertmessage}
\usepackage{amsmath}
\usepackage[colorlinks,linkcolor=blue,citecolor=blue,urlcolor=blue]{hyperref}
\newcommand{\blacktext}[1]{#1}
\usepackage{orcidlink}
\usepackage{xr}
\usepackage{lineno}
\let\oldalign\align
\let\oldendalign\endalign
\renewenvironment{align}{\linenomathNonumbers\oldalign}{\oldendalign\endlinenomath}
\begin{document}
	\title{Fractional Chern Insulators Transition in Non-ideal Flat Bands of Twisted Mono-bilayer Graphene}

	\affiliation{
		State Key Laboratory of Semiconductor Physics and Chip Technologies, Institute of Semiconductors, Chinese Academy of Sciences, Beijing 100083, China
	}
	\affiliation{
		Center for Quantum Matter, Zhejiang University, Hangzhou 310027, China
	}
	\affiliation{
		College of Materials Science and Opto-electronic Technology, University of Chinese Academy of Sciences, Beijing 100049, China
	}

	\author{Moru Song\orcidlink{0009-0003-4842-6959}}
	\affiliation{
		State Key Laboratory of Semiconductor Physics and Chip Technologies, Institute of Semiconductors, Chinese Academy of Sciences, Beijing 100083, China
	}
	\affiliation{
		Center for Quantum Matter, Zhejiang University, Hangzhou 310027, China
	}

	\affiliation{
		College of Materials Science and Opto-electronic Technology, University of Chinese Academy of Sciences, Beijing 100049, China
	}
	
	\author{Kai Chang\orcidlink{0000-0002-4609-8061}}
	\email{kchang@zju.edu.cn}
	\affiliation{
		Center for Quantum Matter, Zhejiang University, Hangzhou 310027, China
	}
	
	\date{\today}
	
	\begin{abstract}
		Fractional Chern insulators (FCIs) in ideal flat bands with Chern number $C$ are commonly understood as color-entangled states constructed from $C$ copies of the lowest Landau level. In realistic moir\'e systems, however, the band geometry is generally non-ideal, and the mechanism that stabilizes such FCIs remains unclear. Using twisted monolayer-bilayer graphene as a platform, we find two FCIs separated by a topological transition that occurs in a regime signaled by a local geometric instability of the Bloch states.
		Below the transition, the target $C=2$ conduction band is geometrically stable, and the resulting fractional phase is naturally described by the Halperin-$(112)$ state. Above the transition, the system becomes geometrically unstable and enters a Laughlin-$1/3$ phase within the same target $C=2$ manifold, which persists even as standard quantum-geometry indicators degrade further. \blacktext{We attribute this Laughlin-$1/3$ phase to a hidden near-ideal $C=1$ component of the non-ideal $C=2$ Bloch states that becomes relevant under interactions, while its strongly non-ideal partner becomes irrelevant.} We support this picture by applying a weak perpendicular magnetic field that acts as a ``color separator,'' directly visualizing the ideal subcomponent at the single-particle level. Together, these results clarify how non-ideal flat bands can stabilize FCIs, greatly expanding their parameter range and sharpening the role of quantum geometry in strongly correlated topological phases.

	\end{abstract}
	
	\maketitle
	
		\textit{Introduction.---}Fractional Chern insulators (FCIs) serve as lattice analogs of fractional quantum Hall (FQH) states and have been experimentally observed in flat bands with $|C| = 1$ \cite{Regnault2011FCI,Bernevig2012EmergentTLS,Liu2022,Xie2021TBGFCI,Cai2023,RMLFCI2024Julong}. Unlike conventional FQH phases in partially filled Landau levels, FCIs can also arise in nearly flat bands with higher Chern numbers $|C|>1$ \cite{Liu2012,WangJie2022CTMG,WangJie2023,DongJunkai2023ColorFCI}, where the ground states often resemble multicomponent Halperin wavefunctions \cite{Halperin1983,LiuZhao2019H221,Regnault2019H_Linterference,WangJie2022CTMG,DongJunkai2023ColorFCI,Ma2024checkerboardFCI} obtained by decomposing the original band into $C$ effective lowest Landau levels (LLLs) or ``colors'' \cite{Tarnopolsky2019CTBG,WangJie2021LL,Qi2012NematicFCI}. This mapping is exact under the ideal flat-band (quantum-geometric trace) condition, which confines vortices bound to electrons within the band subspace and yields purely holomorphic Bloch functions--mirroring the anticlockwise LLL orbitals--so that FCI stability can be predicted from single-particle quantum geometry alone \cite{Ledwith2023Vortexability,Fujimoto2025}. In realistic non-ideal bands beyond LLL, however, vortex attachment becomes nontrivial, Bloch functions acquire anti-holomorphic parts \cite{Ledwith2023Vortexability,Fujimoto2025,Shavit2024QFTFCI}, and the simple color-layer mapping may fail--making explicit many-body calculations essential, as universal single-particle stability criteria remain unknown. Yet, which FCIs survive, and the reason why they can be stabilized in such non-ideal flat bands \cite{Song2026,XiaoDi2026Stability} with $|C|\neq 1$ is still unknown.
	
	\begin{figure*}
		\centering
		\includegraphics[width=\linewidth]{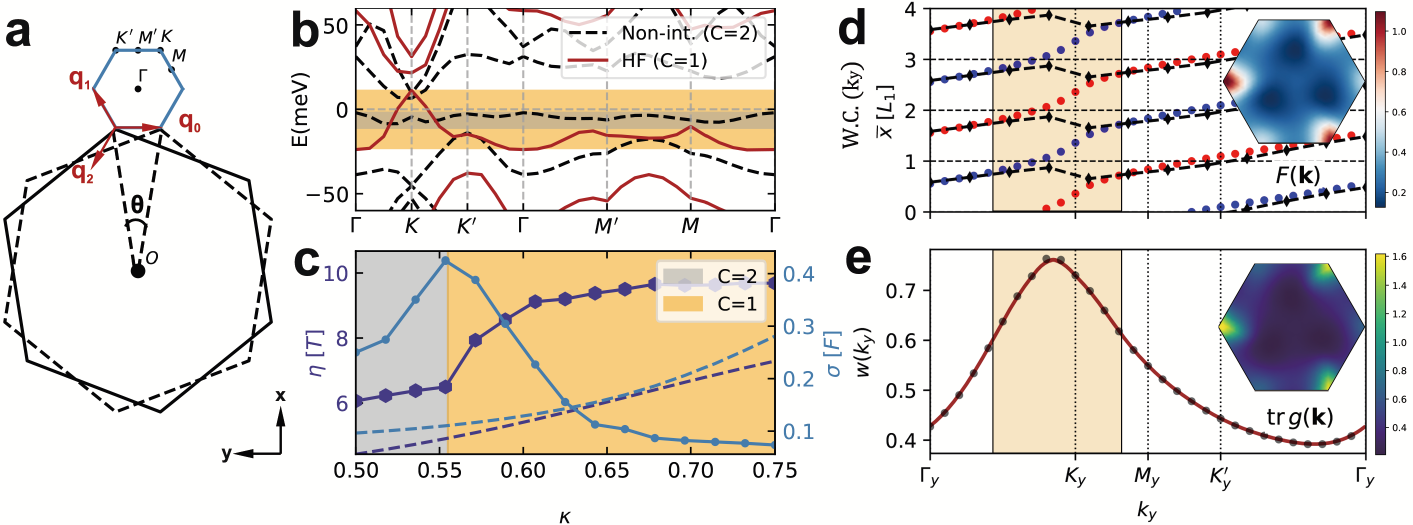}
		\caption{
			    (a) Moir\'e Brillouin zone of tMBG. 
				(b) Non-interacting bands and HF level bands (dashed) of $\xi=1$,$s=\uparrow$ based on CM at $\kappa=0.62$, with target conduction bands shaded in gray and orange. 
				(c) Variation of violation of trace condition (blue) and berry curvature uniformity (orange) in terms of $\kappa$ for non-interacting (solid) and HF cases (dashed).
				(d) Evolution of Wannier centers as a function of $k_y$ for noninteracting (dotted) and HF (dashed black) cases. Blue and red dots denote the two alternating hWF branches of the $C=2$ band, whose centers exhibit distinct winding patterns as a function of momentum \cite{Qi2012NematicFCI}. The $y$-axis spans four unit cells; $L_1$ denotes the length of the moir\'e unit-cell vector. Inset: berry curvature distribution of non-interacting case.
				(e) The spatial extension of hwF $w(k_y)$ evolution with $k_y$ of $C=2$ non-interacted band. The shaded box represents region where the berry curvature concentrate Inset: Metric trace distribution  of non-interacting case.}
		\label{fig:bandstructure}
	\end{figure*}
	
		In this Letter, we use twisted monolayer-bilayer graphene (tMBG) \cite{Bistritzer2011,Rademaker2020TMBG,Ma2021TMBG} to demonstrate a transition between two FCIs in a non-ideal flat band, driven by a geometric instability of the Bloch wave functions. As the parameter $\kappa$ is tuned in the $C=2$ conduction flat bands of the continuum model (CM), the single-particle quantum geometry degrades and an interaction-induced Chern-number reconstruction appears at large $\kappa$. Our exact diagonalization (ED) calculation is performed within a single-band projection onto the target non-interacting $C=2$ conduction band and finds two FCIs separated by a transition at $\kappa_c=0.55$, while Hartree-Fock (HF) at integer filling is used only as a band-theory test of the Bloch-state reconstruction tendency. The transition occurs where hybrid-Wannier analysis of the non-interacting band reveals a local geometric instability.
		\blacktext{For $\kappa<\kappa_c$, the geometrically stable $C=2$ band remains close to an ideal construction with two approximately ideal components entangled through real-space translations.} Interactions then stabilize a Halperin-(112)-like phase with non-chiral edge modes \cite{Wen2004}. 
		For $\kappa>\kappa_c$, while the target bands remain $C=2$ without interaction, the wavefunction becomes geometrically unstable, and the system enters a Laughlin-type FCI with an enhanced deviation from the ideal limit. \blacktext{In this regime, the two components can no longer be viewed as colors entangled through real-space translations. Their relation is instead fixed by the hidden wave-function structure of the target band: interactions make the hidden near-ideal $C=1$ component relevant to the Laughlin-type FCI and its strongly non-ideal partner irrelevant, while the weak perpendicular magnetic field introduced below separates out the same ideal component as a single-particle probe and thereby directly supports this picture} \cite{Xie2021TBGFCI,Bistritzer2011Butterfly,Hejazi2019,Rademaker2020TMBG,Sheffer2021BUtterflyCode}. Such a mechanism for stabilizing FCIs in non-ideal flat bands should extend naturally to higher-Chern bands, broadening the known conditions under which FCIs can emerge.
	
	\textit{Quantum geometry indicators.---}For generic (non-ideal) flat bands, departures from the ideal LLL geometry can be captured by two geometric measures \cite{Xie2021TBGFCI,Ledwith2023Vortexability}: the trace-condition violation $\eta(\mathbf k)=\mathrm{Tr}\,g(\mathbf k)-|F(\mathbf k)|$ and the uniformity of Berry curvature quantified by its standard deviation $\sigma[F(\mathbf k)]$. Here $g_{\mu\nu}(\mathbf k)=\mathrm{Re}\,T_{\mu\nu}$ is the quantum metric and $F(\mathbf k)=-2\,\mathrm{Im}\,T_{xy}$ the Berry curvature, with $T_{\mu\nu}=\langle \partial_{k^\mu}u_{\mathbf k}|(1-|u_{\mathbf k}\rangle\langle u_{\mathbf k}|)|\partial_{k^\nu}u_{\mathbf k}\rangle$ the quantum geometric tensor (or Fubini--Study tensor). These quantities provide useful single-particle criteria for how close a band is to the ideal flat-band limit admitting a $|C|$-component LLL-like description.

		 \textit{Continuum model of tMBG.---}As illustrate in Fig. \ref{fig:bandstructure} (a), tMBG consists of a monolayer stacked atop a Bernal bilayer with a small twist angle $\theta $, giving rise to a moir\'e superlattice and mini Brillouin zone.  The CM of tMBG gives \cite{Bistritzer2011,Rademaker2020TMBG,Ma2021TMBG},
	\begin{align}
		H_0 (\mathbf{k}) =
		\begin{bmatrix}
			h_\theta(\mathbf{k}) & T^\dagger(\mathbf{r}) & 0 \\
			T(\mathbf{r}) & h_\theta(\mathbf{k}) & W^\dagger(\mathbf{k}) \\
			0 & W(\mathbf{k}) & h_\theta(\mathbf{k})
		\end{bmatrix},
		\label{eq:tMBG_CM}
	\end{align}
	where $\mathbf{k}$ is limited in the first Brillouin zone (1BZ), $h_\theta(\mathbf{k}) = -\hbar v_0 \xi (\sigma_x k_1 + \sigma_y k_2)$ is the Dirac Hamiltonian with $(k_1, k_2) = (k_x - K^\xi_x, k_y - K^\xi_y) R^T(\theta)$, $\xi = \pm1$ the valley index, $K^\xi_x,K^\xi_y$ represent the K point of graphene 1BZ at valley $\xi$, and $R(\theta)$ the rotation matrix. The standard Slonczewski-Weiss-McClure form \cite{McCann_2013} of AB-stacked interlayer coupling is
	$
		W(\mathbf{k}) =
		\begin{bmatrix} 
			\hbar v_4 k_+ & \gamma_1 \\
			\hbar v_3 k_- & \hbar v_4 k_+
		\end{bmatrix},
		\label{eq:Gtheta}
	$
	with $k_\pm = \xi k_1 \pm i k_2$ and $\hbar v_\mu = \frac{\sqrt{3} a}{2} \gamma_\mu$. The lattice constant is $a = 2.46\,\mathrm{\AA}$, and hopping parameters are $\gamma_0 = -2.61\,\mathrm{eV}$ and $\gamma_1 = 0.361\,\mathrm{eV}$. In contrast to chiral twisted multilayer graphene model \cite{WangJie2022CTMG,Ledwith2023Vortexability}, where $\gamma_3=\gamma_4=0$, here we have realistic $\gamma_3 = 0.283\,\mathrm{eV}$, $\gamma_4 = 0.140\,\mathrm{eV}$ \cite{Ma2021TMBG}. 
	
		The moir\'e interlayer tunneling $ T(\mathbf{r})=\sum_{s=0}^2 \mathcal{T}_s e^{i\mathbf{q_s\cdot r}}$ is encoded with lowest harmonica components $ \mathcal{T}_s = w_\text{AA} \sigma_0 + w_\text{AB} \left[ \cos(s \xi \varphi) \sigma_x + \sin(s \xi \varphi) \sigma_y \right]$, where $\varphi = 2\pi/3$, $\xi = \pm1$ is the valley index, and $s = 0,1,2$ labels the three $C_3$-related  momentum transfers $\mathbf{q}_s = R^s(\varphi) \cdot \mathbf{q}_0$ with $\mathbf{q}_0 = \frac{8\pi}{3a}\sin(\theta/2)[0,-1]^T$. The tunneling strength ratio is defined as $\kappa = w_\text{AA}/w_\text{AB}$ accounting for the lattice relaxation effect, which usually regard as an experimental parameter that positive related to $\eta[T]$. Furthermore, the external bias $V_\text{pot}=50~meV$ and a sublattice-staggered potential $M=30~meV$  simulating substrate effects within the mono-layer, are added independently. 
	
	\begin{figure*}
		\centering
		\includegraphics[width=0.85\linewidth]{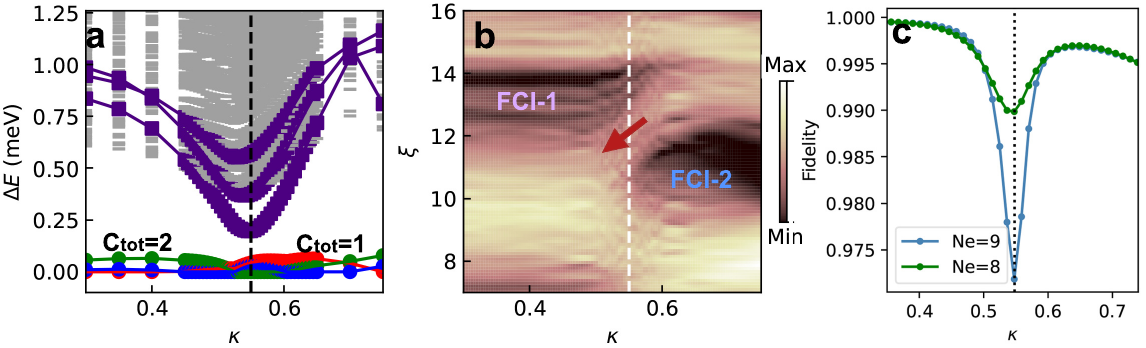}
			\caption{ED calculation on $N_{orb}=4\times 6$ momentum mesh with $N_e=8$ electrons. FCI-1 and FCI-2 stand for FCIs with $C=2/3$ and $C=1/3$. (a) Variation of ED energy levels with $\kappa$, where $\Delta E$ measures relative ground states. Degenerated 3-fold ground states are labeled with red, green and blue. The lowest 3-excited states with same total momentum with respect to ground states are labeled in purple. (b) Variation of PES density of states with $\kappa$, where $e^{-\xi/2}$ is eigenvalue of reduced density matrix $\rho_A$, which is obtained by bipartiting system into $N_A=3,N_B=5$ electrons, $1520$ states below FCI-1 gap while $1088$ sates below FCI-2 gap (see \cite{SI} for details). (c) Density matrix fidelity of ground states for $8$ and $9$ electrons \cite{SI}. }
		\label{fig:edtransition}
	\end{figure*}

		\textit{Band structure signature.---}As a initial test of interaction effect, we begin by performing non-interacting and HF calculation at integer filling $\nu = 1$ of $C=2$ conduction band (See End Matter for projected interaction Hamiltonian).
		The resulting non-interacting band structure at $\kappa = 0.62$ is shown in Fig.~\ref{fig:bandstructure}(b), exhibiting a narrow bandwidth of approximately $9\,\mathrm{meV}$ and a total Chern number $C = 2$  \cite{Rademaker2020TMBG}, with both berry curvature and metric trace concentrate near $K$ points [Inset of Fig. \ref{fig:bandstructure}(d, e)]. However, upon introducing interactions at the $\nu=1$ filling HF level, the target flat band's bandwidth increases significantly to $32\,\mathrm{meV}$, which can be understood as quantum metric-induced hole dispersion following a particle-hole transformation of the fermion operators in the interaction Hamiltonian \cite{Abouelkomsan2023} , and the Chern number spontaneously reduces to $C = 1$. 
	
		Figure~\ref{fig:bandstructure}(c) shows the evolution of two global band-geometric measures, the trace-condition deviation $\eta[T]$ and the Berry-curvature fluctuation $\sigma[F(\mathbf{k})]$, as functions of $\kappa$. 
		In the non-interacting case, where the band remains $|C|=2$ throughout, both $\eta$ and $\sigma$ increase monotonically with $\kappa$, indicating a progressive departure from the ideal flat-band limit. 
		By contrast, at the Hartree--Fock level, $\sigma$ develops a pronounced peak at $\kappa_c=0.55$, accompanied by a discontinuous jump in $\eta$. 
		This $\kappa_c$ coincides with the boundary where the HF-reconstructed band changes from $|C|=2$ to $|C|=1$. 
		For $\kappa>\kappa_c$, the reduced $\sigma$ suggests improved Berry-curvature uniformity, whereas the increased $\eta$ signals a further deterioration of the trace condition. 
		Although the HF Chern-number change is consistent with the FCI transition, as shown in Fig. \ref{fig:bandstructure}(c),  the HF-renormalized quantum geometry---particularly the worsened trace condition---does not by itself account for the stabilization of the Laughlin-like FCI \footnote{Our result contrasts with theoretical predictions for FCIs stabilized in rhombohedral multilayer graphene via pre--Hartree--Fock optimization of the single-particle basis [6, 53, 54, 55].}.
	
		\textit{Geometric instability.---} To locate how the Bloch wave functions reconstruct across the transition, we analyze the bands in the hybrid-Wannier-function (hWF) representation, which keeps one crystallographic direction (here $y$) in momentum space while Wannier-localizing the orthogonal direction $x$, providing a quasi-local probe of single-particle wave-function geometry \cite{SI,Vanderbilt1997,Qi2012NematicFCI,Taherinejad2014WCC}. In this sense, the concentrated Berry curvature and the broadened hWFs are two signatures of the same rapidly varying Bloch states in momentum and real space, respectively.
	
		In this mixed basis, the Wannier center $\bar{x}(k_y)$ traces a flow whose winding counts the Chern number, while its spatial extension $w(k_y)$ directly related to the quantum metric (see End Matter and Supplementary Material \cite{SI}). Thus the hWF representation offers a spatially resolved link between topology--via the center's winding--and geometry--via the shape and thickness of the flow.	
		Figure~\ref{fig:bandstructure}(d) shows the center flow for the noninteracting $C=2$ band (dotted) as alternating hWF branches \cite{Qi2012NematicFCI}. \blacktext{These alternating hWF branches follow from lattice-translation symmetry and single-particle momentum labeling and represent the same $C=2$ Bloch band.} Near $k_y=K_y$, the two branches bend toward each other, which can be understood as effect of concentrated berry-curvature near $K$ (see End Matter). Consistently, Fig.~\ref{fig:bandstructure}(e) reveals a marked increase of $w(k_y)$ in the same window, reflecting an amplified spatial extension of hWF. Curvature concentration draws the trajectories together in momentum space, while metric enhancement spatially extend the hWFs and increases their mutual overlap.
	
		Including interactions, the HF result shown as a dashed black curve in Fig.~\ref{fig:bandstructure}(d) magnifies this geometrically seeded tendency: nearby branches merge within the $K_y$ neighborhood, changing the net winding from two to one and realizing the $C=2\!\to\!1$ jump. \blacktext{The HF $C=2\!\to\!1$ transition proceeds through gap closing and reopening with a nearby band, accompanied by hWF reconnection, band-character transfer, and Berry-curvature sign reversal near $K$.} At the critical point $\kappa_c=0.55$, the center flow develops a momentum-space ambiguity: integrating from $k_y=0$ upward yields $C=2$, whereas integrating downward yields $C=1$, signaling imminent reconnection. These observations define \textit{geometric instability} as rapid variations in the hWF center and spatial extension, which allow interactions to reorganize the hWF flow and its topological response.

		\textit{Topological Transition of FCIs.---}Since interactions can induce a Chern-number transition at integer filling in a full-band calculation, it is natural to extend a similar analysis to fractional fillings within a single-band projection onto the non-interacting $C=2$ conduction band manifold without band mixing with remote band. \blacktext{ED is projected onto the target $C=2$ conduction band.} As shown in Fig.~\ref{fig:edtransition}(a), the neutral excitation mode softens near $\kappa_c$, providing the primary many-body evidence for a transition consistent with the single-particle reconstruction revealed by the HF/hWF analysis.
		The transition is further supported by the particle entanglement spectrum (PES) \cite{LiHaldane2008,Bernevig2008Jack,Regnault2015PES} and the many-body Chern number (MBC). In Fig.~\ref{fig:edtransition}(b), the PES exhibits a clear gap whose level counting below the gap follows the $(k,r)$ admissible rule \cite{Regnault2015PES}, meaning that there are at most $k$ particles in any set of $r$ consecutive orbitals. For $k=1$, the counting is given by $D_{(1,r)}(N_\text{orb},N_e)= \frac{N_\text{orb}} {N_e} {{N_\text{orb}-(r-1)N_e-1}\choose{N_e-1}}$, where $N_\text{orb}=N_x\times N_y$ \cite{SI,Bernevig2012}. In addition, the MBC is obtained by inserting flux through twisted boundary conditions. Under $2\pi$ flux insertion, the degenerate ground states smoothly permute into one another, consistent with a quantized many-body Chern number for the ground-state manifold (Also see Supplementary \cite{SI} for many body berry curvature).
	
		For $\kappa=0.55$, as shown in Fig.~\ref{fig:edtransition}(c), the fidelity of the degenerate ground states displays a pronounced dip increased with system size near the critical point. While this feature are consistent with a continuous transition, they cannot definitively rule out a first-order transition driven by the requisite gap closure for a Chern number change \footnote{More stringent evidence to identify the category of such phase transition may require further studies such as infinite-density matrix renormalization group (iDMRG) or a dedicated critical field theory [56], which is beyond the scope of this work.}, as finite-size systems under flux insertion may exhibit near-critical level crossings \cite{Varney2011TopologicalPT} (see Supplementary Material \cite{SI}, Fig. S11).
	
	For $\kappa<0.55$, we find a threefold degenerate ground-state manifold, which we identify as a Halperin-(112) phase (FCI-1). Each ground state carries Chern number $C=2/3$, and the particle entanglement spectrum (PES) exhibits a clear gap with $D_{(1,2)}(24,3)=1520$ levels below it, consistent with the $(1,2)$ admissible rule. Conventional FCIs realized in higher-Chern-number bands at fillings $\nu = p/(2p|C|+1)$ ($p,|C|\in\mathbb{Z}^+$) are ruled out by their distinct ground-state degeneracy, PES counting \cite{Seidel2008HalperinCounting,Bernevig2012,Seidel2020}, Hall conductance, and fillings \cite{Liu2022,DongJunkai2023ColorFCI,SI}. As shown in the End Matter, the Coulomb ground state of model Halperin-(112) states \cite{McDonald1996,liu2024engineeringfractionalcherninsulators,niu2025quantumanomaloushalleffects} exhibits the same PES counting as obtained here.
	
	We understand the persistence of the Halperin-(112)-like phase only on the mildly non-ideal side of the transition, \blacktext{which is adiabatically connected to an ideal two-color limit with two nearly degenerate LLL-like colors related by real-space translations \cite{DongJunkai2023ColorFCI}. Once this translation relation has been specified, the two colors may be labeled even and odd, with effective fillings $\nu_\mathrm{even}=\nu_\mathrm{odd}=1/3$. This translation-based labeling does not extend to the geometrically unstable regime.} There, trace-condition violation introduces anti-holomorphic weight into the Chern-band representation, while for the Halperin-(112) state at $\nu=1/3$ in the negative Jain sequence $\nu=p/(2p|C|-1)$ with $|C|=2$ and $p=1$, $\det K<0$ implies an indefinite $K$ and a non-chiral edge with corresponding anti-holomorphic factors in standard hierarchy wave-function representations \cite{10.21468/SciPostPhys.8.5.079}. This suggests that non-ideal geometry may be less harmful to the Halperin-(112)-like order than to a fully chiral Laughlin/regular Jain state.
	
		For $\kappa > 0.55$, the PES exhibits a sharp gap with $D_{(1,3)}(24,3)=1088$ states below it, matching the $(1,3)$ admissible rule that characterizes the $\nu = 1/3$ Laughlin-like FCI (FCI-2) and also disfavors a charge density wave interpretation {If a charge-density-wave state were present, the number of states below the PES gap would instead be $3\times \binom{8}{3}=168$.}. Concurrently, the many-body Chern number for each of the nearly threefold degenerate ground states is $C=1/3$, confirming this interpretation \cite{SI}. However, the stabilization of a Laughlin-like FCI in non-ideal $C=2$ flat bands is qualitatively different from that of Halperin-(112) states. To explain the origin of such stabilization mechanism, we propose the following conjecture.
		
	\begin{figure}[b]
		\centering
\includegraphics[width=\linewidth]{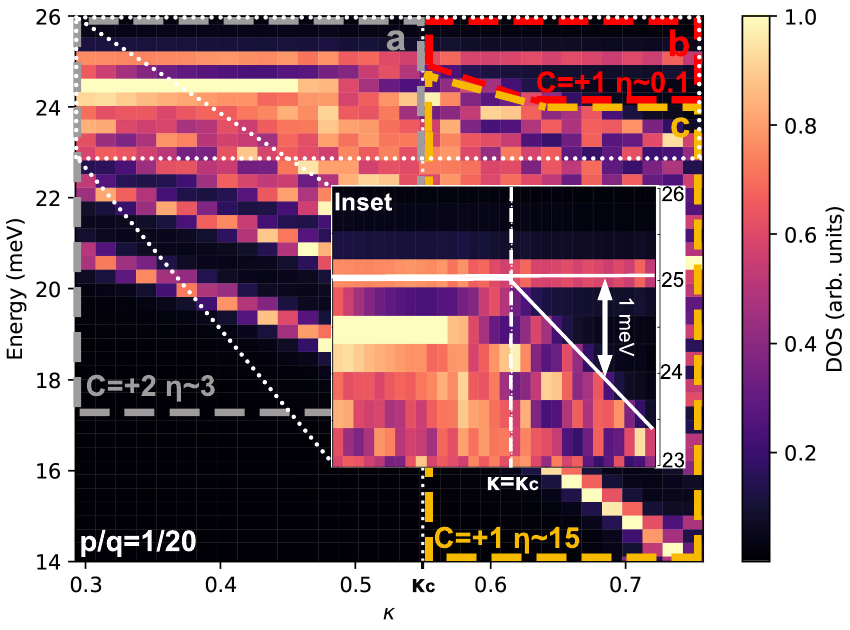}
		\caption{Density of states (DOS) of the target flat bands versus $\kappa$ under weak field $p/q=1/20$. For $\kappa<\kappa_c=0.55$, region \textbf{a} is a connected $C=+2$ manifold with $\eta\sim3$. For $\kappa>\kappa_c$, it separates into a near-ideal $C=+1$ branch in region \textbf{b} with $\eta\sim0.1$ and a strongly non-ideal $C=+1$ branch in region \textbf{c} with $\eta\sim15$, separated by an internal gap of $\sim1\,\mathrm{meV}$. Inset: magnified 23--26 meV window, with white lines as guides to the eye.
		}
		\label{fig:weakmagfiled}
		\vspace{-1.5em}
	\end{figure}
	
		\textit{Color separation conjecture.---}Motivated by ideal-flat-band color decompositions \cite{WangJie2022CTMG,WangJie2023,DongJunkai2023ColorFCI}, we write a non-ideal flat-band Bloch state as
	\begin{align}
		u_{\mathbf k}(\mathbf r)=\sum_\beta {D}_{\mathbf {k}\beta}(\mathbf r)\,v^\beta_{\mathbf k}(\mathbf r),
		\label{eq:NonidealDecomposition}
	\end{align}
			where $u_{\mathbf k}(\mathbf r)$ is the Bloch-periodic part of the target-band wave function, $\mathbf k$ is crystal momentum, $\mathbf r$ is position, $\beta$ labels the effective color components $v^\beta_{\mathbf k}(\mathbf r)$, and ${D}_{\mathbf{k}\beta}(\mathbf r)$ is the corresponding momentum- and position-dependent weighting function. For the stable ideal limit below, $v_{\mathbf k}(\mathbf r)$ denotes a reference color component and $\mathbf a_\beta$ is the real-space lattice-translation vector associated with color $\beta$. A formal decomposition of a generic non-ideal band can be nonunique. \blacktext{In the geometrically stable ideal limit, the two components are explicitly related by real-space translations, $v^\beta_{\mathbf k}(\mathbf r)=v_{\mathbf k}(\mathbf r+\mathbf a_\beta)$, and are entangled through this relation. In the geometrically unstable target band, the components can no longer be viewed as two colors entangled through real-space translations. Once the target band and its wave-function structure are specified, the underlying decomposition is fixed, although it remains hidden within the same zero-field $C=2$ Bloch states. The target-band form factors encode this hidden connection, and inter-$\mathbf k$ mixing induced by interactions or a weak magnetic field makes it operative and numerically resolvable. Equation~\eqref{eq:NonidealDecomposition} describes both regimes and provides a generalization of the ideal color-decomposition picture to non-ideal flat bands.} Usually, the number of colors is expected to be $|C|$. The weighting function ${D}_{\mathbf{k}\beta}(\mathbf r)$ encodes how the Bloch-periodic wave function distributes its real-space weight among color sectors, connecting with the real-space charge distribution. Different components can carry sharply different quantum geometry, quantified by the component-resolved trace-condition violation $\eta^\beta[v^\beta_{\mathbf k}]$, namely the trace-condition violation evaluated for component $v^\beta_{\mathbf k}$. \blacktext{Interaction-induced inter-$\mathbf k$ scattering through the same form factors activates this pre-existing color separation. At total target-band filling $\nu=1/3$, the hidden near-ideal component becomes relevant to the Laughlin-like state, while its strongly non-ideal partner becomes irrelevant.} Accordingly, the geometrically stable region ($\kappa<\kappa_c$) remains close to the ideal translation-entangled two-color limit, whereas in the geometrically unstable region ($\kappa>\kappa_c$) the relation between the two components is hidden in the original $C=2$ wave functions.
	
    \begin{figure}
		\centering
		\includegraphics[width=\linewidth]{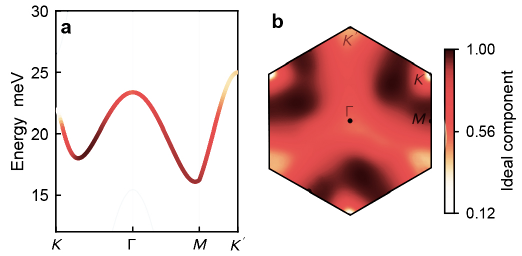}
		\vspace{-0.5em}
		\caption{Momentum-resolved projection of the weak-field-separated near-ideal $C=1$ component onto the zero-field target band at $\kappa=0.7$ and $p/q=1/20$. a, Projected weight on the target band. b, The same normalized weight in the hexagonal moir\'e Brillouin zone.}
		\label{fig:weakfieldcomponenthex}
	\end{figure}

		\textit{Color separation under magnetic field.---}When $\kappa>\kappa_c$, the Bloch wave functions develop a geometric instability near $K$. Ordinary-translation-breaking terms mix momenta through $\langle u_{\mathbf{k+q}}|u_{\mathbf{k}}\rangle \approx 1-\tfrac12 g_{\mu\nu}(\mathbf{k})q^\mu q^\nu$, \blacktext{so the same form factors govern how interaction-induced scattering activates the color separation and how a weak probe directly resolves the same pre-existing components.} We therefore apply a weak perpendicular magnetic field with flux $\phi/\phi_0=p/q=1/20$ per moir\'e unit cell. Figure~\ref{fig:weakmagfiled} shows that for $\kappa<\kappa_c$ the target manifold remains a connected $C=+2$ band with $\eta\sim3$, whereas for $\kappa>\kappa_c$ it separates into a near-ideal $C=+1$ branch with $\eta\sim0.1$ and a strongly non-ideal $C=+1$ partner with $\eta\sim15$, separated by an internal gap of $\sim1\,\mathrm{meV}$. \blacktext{The zero-field FCI is established by ED, while the weak field directly resolves the pre-existing hidden near-ideal $C=1$ component. The splitting begins at $\kappa_c$, coincident with the zero-field FCI transition.} An independent weak-field calculation reproduces the same spectrum and splitting (see Supplemental Material). \blacktext{A direct comparison of the ideal component separated out by the weak field with the zero-field target-band states gives a $91.9\%$ overlap, demonstrating that this ideal component is embedded in the original $C=2$ target-band wave functions. Its weight remains continuously distributed throughout the moir\'e Brillouin zone but is strongly nonuniform: it is only about $0.12$ near the symmetry-related $K$ points, remains small at the other high-symmetry points, and is concentrated mainly at generic momenta where the quantum metric and Berry curvature are not concentrated [see the insets of Fig.~\ref{fig:bandstructure}(d,e)].} \blacktext{The code used to identify and characterize the ideal component through this weak-field magnetic probe is publicly available in Ref.~\cite{SongChang2026MoireFCICode}.}

	    
	    The Hofstadter butterfly and trace conditions of magnetic subbands are discussed in End Matter [Fig.~\ref{fig:butterfly}], consistent with recent observations of FCI signatures in $C=2$ tMBG bands \cite{tMBG_exp2025}.
	
	\textit{Conclusion.---}We identify a local geometric instability of the Bloch states, revealed by HF and hWF analyses, that signals the FCI transition regime in non-ideal $|C|>1$ flat bands. Within a single-band projection onto the non-interacting $C=2$ conduction band, ED reveals a many-body transition between two distinct FCIs near $\kappa_c$. \blacktext{The $91.9\%$ overlap with the zero-field target band shows that the weak-field response resolves a pre-existing hidden near-ideal $C=1$ component embedded in the original $C=2$ Bloch states. This component is relevant to the Laughlin-like FCI and remains continuously distributed across the Brillouin zone, although its weight is strongly suppressed at the high-symmetry points.} Related component-selection physics in other non-ideal FCI-supporting flat bands is left for future work \cite{Song2026,Simon2015,liu2025topologicalorderbandtopology}. These results position tMBG as a controlled platform to probe the interplay between interaction and geometry and to engineer topological phase transitions in moir\'e systems.
	
	\textit{Note added.---}During preparation, a transition between Halperin and abnormal Laughlin states was also found in twisted double bilayer graphene \cite{liu2025topologicalorderbandtopology}. 
	
	\begin{acknowledgments}
		We acknowledge Zhao Liu for helpful discussion. This work is supported by the National Natural Science Foundation of China (NSFC; Grants No. 12488101, No. 12574058, and No. 92265203), the Strategic Priority Research Program of the Chinese Academy of Sciences (Grants No. XDB0460000 and No. XDB28000000), and the Quantum Science and Technology-National Science and Technology Major Project (Grants No. 2024ZD0300104 and No. 2021ZD0302600).
	\end{acknowledgments}
	\bibliography{ColorFCI.bib}
	
	\appendix*
	\setcounter{equation}{0}
	\onecolumngrid
	\section*{End Matter}
	\twocolumngrid
	
	\textit{Projected Interaction Hamiltonian.---}Together with the interaction terms, the full target-band interaction Hamiltonian---obtained after diagonalizing the CM---can be written as \cite{Zaletel2020, Kwan2023HF},
	\begin{align}
		H=\sum_{nk} E_{n\mathbf{k}} \hat n_{n,\mathbf{k}} + \frac{1}{2A}\sum_\mathbf{q} V(\mathbf{q}):\hat \rho_\mathbf{q} \hat\rho_\mathbf{-q}:,
		\label{eq: InteractionHamitonian}
	\end{align}
	where $ \hat n_{n,\mathbf{k}}= \sum_{s,\xi}c_{n,\mathbf{k},s,\xi}^\dagger c_{n,\mathbf{k},s,\xi}$ are particle number operator of non-interacting bands $E_{n\mathbf{k}}$,. $\hat\rho_\mathbf{q}=\sum_{\mathbf{k},m,n,\sigma,\xi,\sigma',\xi'} \braket{u_{\mathbf{k+q},m,s,\xi}|u_{\mathbf{k},n,s',\xi'}} c_\mathbf{k+q,m,s,\xi}^\dagger c_\mathbf{k,n,s',\xi'}$ is the density operator, $A$ is the area of sample, $s=\uparrow,\downarrow$ is spin index, and $V(\mathbf{q}) = \frac{e^2}{2 \varepsilon q} \tanh(d_\text{sc} q / 2)$, where $\varepsilon\approx 4\varepsilon_0$ is the dielectric constant of the material, $d_\text{sc}=10~nm$ is considered as the distance towards the gate. Here, we consider a model with two spin and two valley flavors, including three active conduction and three valence bands per flavor, along with remote band renormalization effects in average scheme \cite{Kwan2023HF,Yu2024}. Within this setup \footnote{For technique details on HF calculation with similar set up, we recommend readers also see [53]}, the self-consistent HF ground state is found to be spin-valley polarized \cite{SI}.

	\begin{figure}
		\centering
		\includegraphics[width=\linewidth]{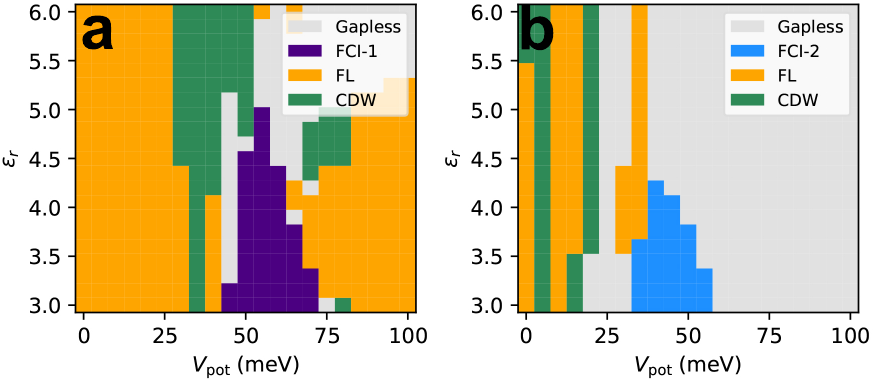}
		\caption{(a, b) The phase diagram at $\kappa=0.5$ and $\kappa=0.7$ in terms of external bias $V_\mathrm{pot}$ and relative dielectric constant $\varepsilon_r$.}
		\label{fig:phasediagram}
	\end{figure}
	
	In the main text, we utilize ED to compute the many-body ground states by projecting Eq. \eqref{eq: InteractionHamitonian} onto non-interacting  $C=2$ conduction band,
	and we use the PES to identify their topological character. We also compute the many-body Chern number (MBC) of the ground-state manifold \cite{SI,Okamoto2022}.
	The PES is obtained by tracing out a subset of  particles from the degenerate ground-state manifold and examining the eigenvalue spectrum of the resulting reduced density matrix $\rho_A=\text{Tr}_B\rho$, and $\xi = -2\text{log} \rho_A$ with $\rho=\frac{1}{D}\sum_D\ket{\Omega_D}\bra{\Omega_D}$ is the ground state density matrix. When the spectrum is gapped, the number of low-energy states below the gap provides a robust fingerprint of the topological order, as it precisely follows the expected counting rules characteristic of specific topological phases \cite{LiHaldane2008,Bernevig2008Jack,Regnault2015PES}.
	
	\textit{Phase diagram.---}Importantly, the emergent of FCIs needs specific conditions in tMBG. As depicted in Fig. \ref{fig:phasediagram}, we present phase diagrams for both types of FCI phases at $\kappa = 0.5$ and $\kappa = 0.7$, evaluated over varying interaction strength based on dielectric constant $\varepsilon_r$ and perpendicular electric basis $V_\text{pot}$. In both cases, stronger interactions and an fine-tune bias $V_\mathrm{pot}\sim50\,\mathrm{meV}$ are required to stabilize the FCI phases. Notably, FCI-1 favors a larger range of area of phase diagram. While competing phases such as Fermi liquid (FL), charge density wave (CDW), and other gapless states can be distinguished in ED by their energy gaps, structural factors and ground-state degeneracies \cite{SI}.

\begin{figure}
		\centering
		\includegraphics[width=1\linewidth]{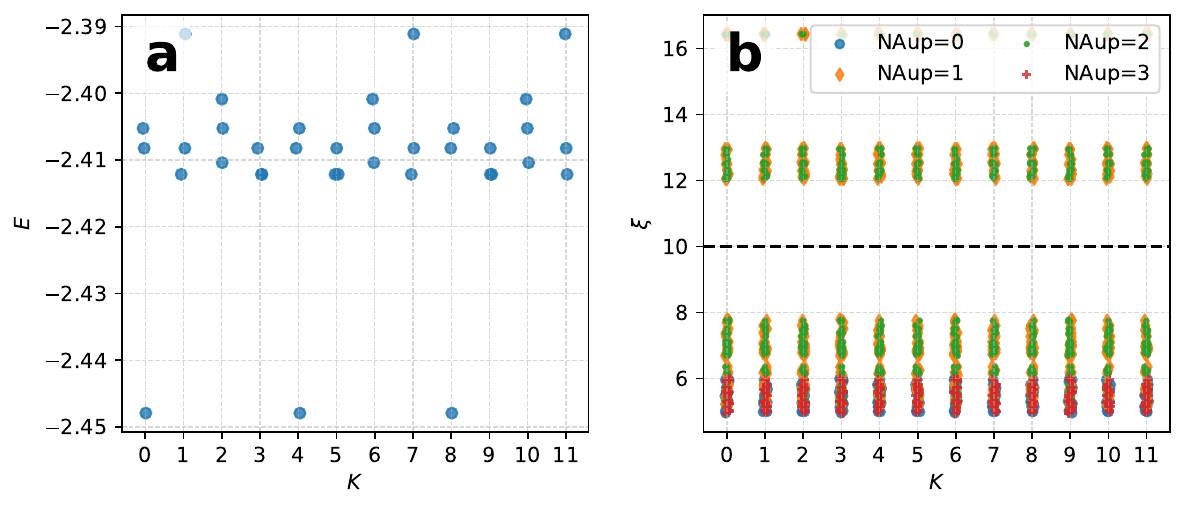}
		\caption{(a) Energy spectrum of Halperin-(112) FQH states under coulomb interaction with $N_A=8,N_\phi=12,S_z=0$. (b) Corresponding PES which matches (1,2)-counting with $N_A=3$ for spin up with $S_z=0,1,2,3$, where $1520$ states below the dashed line.}
		\label{fig:h112energyandpessubplots}
	\end{figure}

	\textit{Halperin-(112) states---} To verify that the FCI-1 phase in our work shares the same topological order as the Halperin-(112) state, we employ the Coulomb Hamiltonian of Ref.~\cite{McDonald1996}, which exhibits nearly unit overlap with the model wave function below,
	\begin{align}
		\Psi(z^\uparrow,z^\downarrow)&=\mathcal{P}_{LLL} \prod_{i<j}(z^{\uparrow*}_i-z^{\uparrow*}_j)\prod_{i<j}(z^{\downarrow*}_i-z^{\downarrow*}_j)\label{eq:H112}\\\nonumber &\times\prod_{i<j}(z^\uparrow_i-z^\uparrow_j)^2 
		\prod_{i<j}(z^\downarrow_i-z^\downarrow_j)^2\prod_{i\neq j}(z_i^\uparrow-z^\downarrow_j)^2.		
	\end{align}
	Here $\mathcal{P}_{LLL}$ projects onto the lowest Landau level, and $z^\uparrow$ ($z^\downarrow$) are the complex coordinates of spin-up (spin-down) electrons, where the spin here indicates layer degree of freedom. Although Eq.~\eqref{eq:H112} is not the standard form of the Halperin-(112) state, its $K$ matrix and charge vector coincide with those of the standard construction \cite{McDonald1996}, implying the same topological order, i.e. $K=K^+-K^-=\begin{bmatrix} 2 & 2\\ 2 & 2 \end{bmatrix}-\begin{bmatrix} 1 & 0\\ 0 & 1 \end{bmatrix}=\begin{bmatrix} 1 & 2\\ 2 & 1 \end{bmatrix}$ \cite{10.21468/SciPostPhys.8.5.079}. In Fig.~\ref{fig:h112energyandpessubplots}(a) we diagonalize the Coulomb Hamiltonian [Eq.~(50) of Ref.~\cite{McDonald1996}] at layer separation $d=0$ with $8$ electrons and $12$ flux quanta, and find a threefold-degenerate ground-state manifold. The PES exhibits the same characteristic counting, with $1520$ levels below the entanglement gap, consistent with the same topological order as in our system. These results support that the FCI-1 phase in our system is Halperin-(112)-like.
	
	\begin{figure*}
		\centering
		\includegraphics[width=\linewidth]{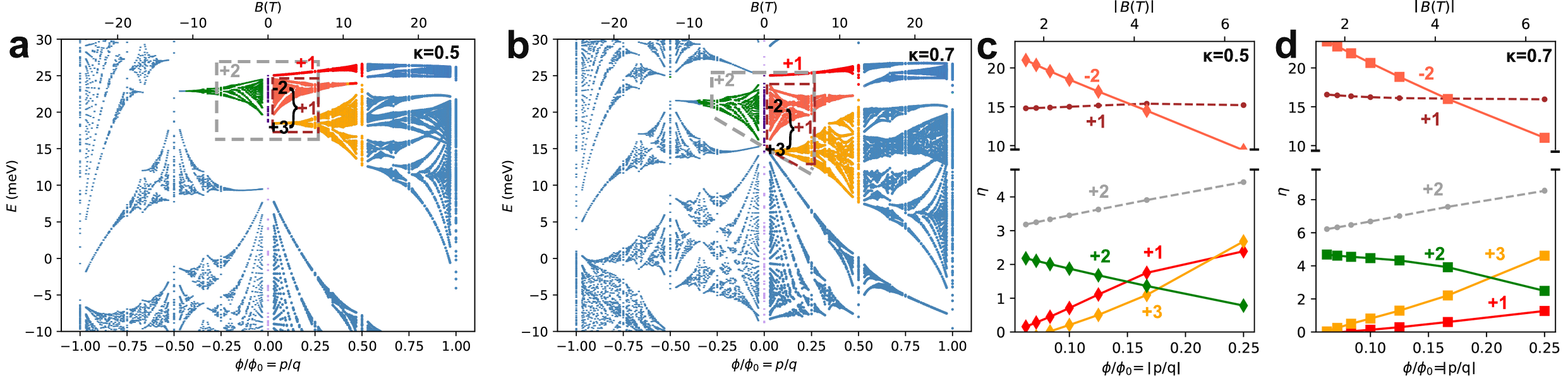}
		\caption{(a,b) Hofstadter Butterfly at $\kappa=0.5$ and $\kappa=0.7$ with $p,q<20$ for $+K$ valley, where Chern numbers are labeled and can be verified from Diophantine equation of Wannier plot \cite{SI}. (c,d) The violation of trace condition variation in terms of magnetic field for magnetic subbands within the box in (a,b) with same color. The dashed line represents $\eta$ for multi-bands }
		\label{fig:butterfly}
	\end{figure*}
	
	\textit{Hybrid Wannier functions and quantum geometry.---}Using single-particle Bloch eigenstates, we construct hybrid Wannier functions (hWFs) that are maximally localized along $\hat x$ at fixed $k_y$
	\cite{SI,Vanderbilt1997,Qi2012NematicFCI,Taherinejad2014WCC}.
			In real space, the hWF is localized along $x$ and extended along $y$, analogous to Landau-gauge Landau levels; sweeping $k_y$ gives its center trajectory $\bar x_n(k_y)$ and width. The center flow and spatial extension characterize this evolution. The slope of the trajectory is set by the Berry curvature,
	$
	\frac{d\,\bar x_n(k_y)}{dk_y}=\frac{L_1}{2\pi}\oint_{\mathrm{BZ}_x}\!dk_x\,F(\mathbf k),
	$
	while the spatial extension of the hWF at fixed $k_y$ is quantified by the metric-trace definition we adopt,
	$
	w(k_y)\sim\sqrt{\,L_1\oint_{\mathrm{BZ}_x}\frac{dk_x}{2\pi}\,\mathrm{Tr}\,g(\mathbf k)\,}\,,
	$
			consistent with using $\mathrm{tr}\,g$ to capture the gauge-invariant spread \cite{Vanderbilt1997}. Thus Landau levels or ideal flat bands have a straight hWF trajectory with constant slope and width; in the lowest Landau level, the \emph{trace condition} $\mathrm{Tr}\,g(\mathbf k)=|F(\mathbf k)|$ matches the slope and width along the path \cite{Ledwith2023Vortexability}. In realistic situations, overly large width makes those momenta less favored by exchange and raises the local Fock energy; this is captured by the exchange-induced (hole) dispersion \cite{Abouelkomsan2023}
	$\tilde\varepsilon_{-\mathbf k}\simeq \sum_{\mathbf q}V(\mathbf q)\,e^{-g_{ab}(\mathbf k)\,q_a q_b},$
			so regions with large $\mathrm{Tr}\,g(\mathbf k)$ are less occupied at HF level, explaining the lifted HF band near $\mathbf K$ and the enhanced dispersion in Fig. \ref{fig:bandstructure} (a). In our case, Berry-curvature hotspots appear as pronounced hWF bends and coincide with width peaks; where two $C=2$ trajectories approach, broader hWFs interfere more strongly and reconnect, providing a transparent picture for the Chern-number-changing transition. \blacktext{The near-ideal-component weight remains continuous across the Brillouin zone but is strongly suppressed at high-symmetry momenta---reaching only about $0.12$ near $K$---and is concentrated mainly at generic momenta where the quantum metric and Berry curvature are not concentrated [see the insets of Fig.~\ref{fig:bandstructure}(d,e)].}

			\textit{Relation with color-entangled ideal flat bands.---}Without geometric instability, a non-ideal flat band may be adiabatically connected to its ideal limit. In an ideal flat band, the Bloch wavefunction can be chosen holomorphic, as for an LLL in a non-uniform magnetic field. \blacktext{Target-band vortexability links this momentum-space ideal geometry to real-space LLL analyticity \cite{Ledwith2023Vortexability}, whose many-body counterpart is the  Jastrow factor $\prod_{i<j}(z_i-z_j)^m$.} Using the same notation as Eq.~\eqref{eq:NonidealDecomposition}, the color-entangled Bloch wavefunction takes the form
	$u_\mathbf{k}(\mathbf{r})=\sum_\beta D_{\mathbf{k}\beta}(\mathbf{r})v^\beta_\mathbf{k}(\mathbf{r}),$
			where $u_\mathbf{k}(\mathbf{r})$ is the periodic part of the flat-band Bloch wavefunction. In the ideal translation-entangled limit, $v^\beta_\mathbf{k}(\mathbf{r})=v_\mathbf{k}(\mathbf{r}+\mathbf{a}_\beta)$ and $D_{\mathbf{k}\beta}(\mathbf{r})=D_\mathbf{k}(\mathbf{r}+\mathbf{a}_\beta)$, with lattice vectors $\mathbf{a}_\beta=\sum_{i=1,2}\beta_i\mathbf{a}_i$ and $\beta_i=0,\dots,C_i-1$ under the gauge choice $C=C_1\times C_2$ \cite{WangJie2022CTMG,WangJie2023,DongJunkai2023ColorFCI}. \blacktext{Real-space translations shift one color component into another in the ideal limit, where an enlarged-cell or reduced-zone $C=1$ sector gives an equivalent representation. By contrast, in the geometrically unstable non-ideal regime, the two components can no longer be viewed as colors entangled by those translations. Their relation is fixed but hidden in the target-band wave functions and form factors, and no reduced-zone construction is introduced. Both limits are described by Eq.~\eqref{eq:NonidealDecomposition}.} Generally, the color-separation effect may stabilize FCIs not only in $|C|>1$ flat bands but also in $|C|=1$ or trivial non-ideal flat bands \cite{liu2025topologicalorderbandtopology,Lin2026ZeroChernFCI} with non-uniform geometries. Developing a systematic understanding of the color-separation effect could significantly enhance our insights into the relationship between quantum geometry and strongly correlated phases.
	
	\textit{Hofstadter butterfly.---}As shown in Fig.~\ref{fig:butterfly}(a,b), we compute the magnetic spectrum of tMBG at the $+K$ valley (for the $-K$ valley, the magnetic field and Chern numbers are reversed). Under positive magnetic field, the original $C = +2$ flat band splits into subbands with Chern numbers $+1$, $-2$, and $+3$. In contrast, under negative magnetic field, the band remains intact with $C = +2$.	
		We further evaluate the trace-condition violation of magnetic subbands. At both $\kappa=0.5$ and $\kappa=0.7$, the separated subbands that require stronger fields than the color-separation regime and have Chern numbers $C=1$ and $C=3$ exhibit nearly ideal trace conditions, with $\eta \sim 0.1$. In contrast, only at $\kappa=0.5$ does the original $C=2$ band attain $\eta \sim 2$ under a weak negative magnetic field. This means stronger negative field can stabilize Laughlin and Halperin-like states generally.

\end{document}


\title{Supplementary Material for Fractional Chern Insulators Transition in Non-ideal Flat Bands of Twisted Mono-bilayer Graphene}

	\affiliation{
		State Key Laboratory of Semiconductor Physics and Chip Technologies, Institute of Semiconductors, Chinese Academy of Sciences, Beijing 100083, China
	}
	\affiliation{
		Center for Quantum Matter, Zhejiang University, Hangzhou 310027, China
	}
	\affiliation{
		College of Materials Science and Opto-electronic Technology, University of Chinese Academy of Sciences, Beijing 100049, China
	}

	\author{Moru Song\orcidlink{0009-0003-4842-6959}}
	\affiliation{
		State Key Laboratory of Semiconductor Physics and Chip Technologies, Institute of Semiconductors, Chinese Academy of Sciences, Beijing 100083, China
	}
	\affiliation{
		Center for Quantum Matter, Zhejiang University, Hangzhou 310027, China
	}
	
	\affiliation{
		College of Materials Science and Opto-electronic Technology, University of Chinese Academy of Sciences, Beijing 100049, China
	}
	
	\author{Kai Chang\orcidlink{0000-0002-4609-8061}}
	\email{kchang@zju.edu.cn}
	\affiliation{
		Center for Quantum Matter, Zhejiang University, Hangzhou 310027, China
	}
	
	\date{\today}
	
	\maketitle
	\onecolumngrid
	\tableofcontents
	\
	\twocolumngrid
\section{Hartree Fock calculation}

In the projected Hartree–Fock (HF) approach, we write the total mean-field Hamiltonian as a sum of single-particle and interaction terms, $\hat{\mathcal{H}}_{\mathrm{HF}} = \hat{\mathcal{H}}_0 + \hat{\mathcal{H}}_{\mathrm{bg}} + \hat{\mathcal{H}}_{\mathrm{int}}[\delta \rho]$. Here $\hat{\mathcal{H}}_0$ is the non-interacting moiré band Hamiltonian (including kinetic energy and any external potential such as interlayer bias), $\hat{\mathcal{H}}_{\mathrm{bg}}$ is the static mean-field due to the \textbf{remote bands} (those outside the active window), and $\hat{\mathcal{H}}_{\mathrm{int}}[\delta \rho]$ is the self-consistent interaction term from the \textbf{active bands}. Following the scheme of \cite{kwan2023HF,Yu2024}, the remote valence bands (fully filled in the non-interacting picture) and remote conduction bands (empty) are treated as an inert background (average scheme \cite{kwan2023HF,Yu2024}): we subtract a reference half-filling for each remote state so that only their deviation (filled $-\,\tfrac{1}{2}$ or empty $-\,\tfrac{1}{2}$) contributes to a background Hartree–Fock potential. This ensures that the high-energy bands do not explicitly enter the variational HF energy, aside from renormalizing the active bands via a fixed one-body potential. We now define the key tensors entering the HF formulation. First, the umklapp matrix $\Gamma$ enforces momentum conservation up to a moiré reciprocal lattice vector $\mathbf{G}$. In an extended plane-wave basis (indexing plane-wave momenta by integers $g,g'$), $\Gamma_{g\,g'}(\mathbf{G}) = \delta_{\mathbf{G}_g,\;\mathbf{G}_{g'}+\mathbf{G}}$. Next, the \textbf{form factor tensor} $\Lambda$ encodes the inter-band overlap of Bloch states upon momentum transfer. In particular, for a given valley $\xi$ and layer $l$, we define $\Lambda^{(\xi,l)}_{m n}(\mathbf{k},\mathbf{q}+\mathbf{G})$ as the overlap between an active Bloch state $|u^{(\xi)}_{n,\mathbf{k}}\rangle$ and another state $|u^{(\xi)}_{m,\mathbf{k}+\mathbf{q}}\rangle$ after a momentum boost by $\mathbf{q}+\mathbf{G}$. Writing $U^{(\xi)}_{n,\mathbf{k}}(g,l,\alpha)$ for the plane-wave expansion coefficient of $|u^{(\xi)}_{n,\mathbf{k}}\rangle$ on layer $l$ and sublattice $\alpha$ (with $g$ labeling the moiré reciprocal lattice vectors in the plane-wave cutoff), we have the explicit form factor:
\begin{align}
	&\Lambda^{(\xi,l)}_{m n}(\mathbf{k},\,\mathbf{q}+\mathbf{G})	\label{Lambda_def}\\\nonumber 
	&= \sum_{g_1,\,g_2,\,\alpha} U^{(\xi)*}_{m,\;\mathbf{k}+\mathbf{q}}\!(g_1,l,\alpha)\;\,\Gamma_{g_1 g_2}(\mathbf{G})\;\,U^{(\xi)}_{n,\;\mathbf{k}}\!(g_2,l,\alpha)~, 
\end{align}
which is an overlap matrix element between Bloch states at $\mathbf{k}$ and $\mathbf{k}+\mathbf{q}$ with a reciprocal shift $\mathbf{G}$. This tensor $\Lambda^{(\xi,l)}_{m n}(\mathbf{k},\mathbf{q}+\mathbf{G})$ carries indices for the initial band $n$, final band $m$, valley $\xi$, layer $l$, and momenta $\mathbf{k}$ (in the moiré Brillouin zone) and $\mathbf{q}+\mathbf{G}$ (the momentum transfer, decomposed into a small momentum $\mathbf{q}$ in the first moiré Brillouin zone plus a reciprocal lattice vector $\mathbf{G}$, i.e. an umklapp component). 
\subsection{Dual gate interaction}
In our main text, we use the two dimensional dual gate interaction,
\begin{gather}
	V_{L,l}(\mathbf{q})=\frac{e^2}{2\varepsilon q A}\tanh(q d_\text{sc}),
	\label{eq:2Dinteraction}
\end{gather}
which is independent of layer. In general, the interaction should layer dependents as given in ref  \cite{kwan2023HF}. In the derivation below, we assume the most general cases and write the layer index $L,l=0,1,2$ explicitly. For a layer dependence interaction, one can use $V_{L,l}(\mathbf{q})$ in \cite{kwan2023HF},
\subsection{Background terms}
Using the form factors, we can express the \textbf{background density matrix} due to remote bands. Let $f_{m,\xi}$ be the occupation of remote band $m$ in valley $\xi$ (so $f_{m,\xi}=1$ for filled remote valence bands, $0$ for empty remote conduction bands). The reference subtraction scheme sets each remote $f_{m,\xi}^{\text{ref}}=\tfrac{1}{2}$, so the net background occupancy is $\delta f_{m,\xi}=f_{m,\xi}-\tfrac{1}{2}=\pm\tfrac{1}{2}$. The Fourier-component of the background charge density on layer $l$ at momentum $\mathbf{G}$ is then 

\vspace{-1em}
\begin{equation}
	\mathcal{N}_{l}^{\mathrm{bg}}(\mathbf{G}) \;=\; \frac{1}{N_{\mathrm{cell}}}\sum_{\xi}\sum_{\substack{m\in \text{remote}}}\sum_{\mathbf{k}} \Big(f_{m,\xi}-\tfrac{1}{2}\Big)\;\Lambda^{(\xi,l)}_{m m}(\mathbf{k},\,\mathbf{G})~,
	\label{Nbg_def}
\end{equation}

where $N_{\mathrm{cell}}$ is the number of moiré unit cells in the periodic HF supercell (so that $\frac{1}{N_{\mathrm{cell}}}\sum_{\mathbf{k}}$ indicates an average over the moiré Brillouin zone). In Eq.~(\ref{Nbg_def}), $\Lambda_{m m}(\mathbf{k},\mathbf{G})$ is the intra-band form factor giving the weight of Bloch state $m$ at wavevector $\mathbf{k}$ on layer $l$ with momentum transfer $\mathbf{G}$; thus $\mathcal{N}_l^{\mathrm{bg}}(\mathbf{G})$ represents the Fourier-space density (relative to the neutralizing reference) contributed by all remote bands on layer $l$. In particular, $\mathcal{N}_l^{\mathrm{bg}}(0)$ yields the total background charge in layer $l$, while nonzero $\mathbf{G}$ components encode the periodic charge modulation from remote-band Bloch states. 

We can now write the HF mean-field Hamiltonian acting within the active subspace. The \textbf{background Hartree–Fock term} $\hat{\mathcal{H}}_{\mathrm{bg}}$ is a one-body operator representing the static Coulomb field of the remote-band charges (both Hartree direct and Fock exchange) felt by the active electrons. It can be expressed as 

\vspace{-1em}
\begin{equation}
	[H_{\mathrm{bg}}]_{n n'}(\mathbf{k}) \;=\; [H_{\mathrm{bg}}^{H}]_{n n'}(\mathbf{k}) + [H_{\mathrm{bg}}^{F}]_{n n'}(\mathbf{k})~, 
	\label{H_bg_sum}
\end{equation}

with separate contributions from the Hartree (H) and Fock (F) parts. In the Hartree (direct) term $H_{\mathrm{bg}}^{H}$, the active electrons on layer $L$ experience an electrostatic potential from the background charge in layer $l$. In Fourier space this is given by the Coulomb interaction $V_{L l}(\mathbf{G})$ between a charge on layer $L$ and a charge on layer $l$ at relative momentum $\mathbf{G}$. Thus, using the background density $\mathcal{N}_l^{\mathrm{bg}}$ defined above, we have 

\vspace{-1em}
\begin{equation}
	[H_{\mathrm{bg}}^{H}]_{n n'}(\mathbf{k}) \;=\; \sum_{L,\,l}\;\sum_{\mathbf{G}} V_{L l}(\mathbf{G})\;\Lambda^{(\xi,L)}_{n n'}(\mathbf{k},\,\mathbf{G})\;\mathcal{N}_{l}^{\mathrm{bg}}(-\mathbf{G})~,
	\label{Hbg_H}
\end{equation}

where $\Lambda^{(\xi,L)}_{n n'}(\mathbf{k},\mathbf{G})$ serves as a form-factor projector of the active $nn'$ state (at valley $\xi$) onto layer $L$ for the plane-wave component $e^{i\mathbf{G}\cdot\mathbf{r}}$, and $\mathcal{N}_{l}^{\mathrm{bg}}(-\mathbf{G})$ is the background charge on layer $l$ with the opposite momentum $-\mathbf{G}$ (ensuring overall momentum conservation in the Hartree term). Equation~(\ref{Hbg_H}) shows that a nonzero $\mathcal{N}_l^{\mathrm{bg}}(\mathbf{G})$ (coming from remote-band charge imbalance) generates a Hartree potential acting on the active bands via their form factors $\Lambda_{nn'}$. Meanwhile, the \textbf{Fock term} $H_{\mathrm{bg}}^{F}$ represents the exchange self-energy contributed by the filled remote bands. An active electron in band $n'$ can virtually exchange with an electron in a remote band $m$ (valence) or a hole in a remote band $m$ (conduction) of the same valley, resulting in a correction to the one-body energy of the active state $n'$. Using the form factors to handle wavefunction overlaps, we write the background Fock contribution as 
\begin{align}
	&[H_{\mathrm{bg}}^{F}]_{n n'}(\mathbf{k}) = -\sum_{L,\,l}\sum_{\mathbf{q},\mathbf{G}} \frac{V_{L l}(\mathbf{q}+\mathbf{G})}{N_{\mathrm{cell}}} \times\nonumber\\
	&\sum_{\substack{m\in \text{remote}}}\delta f_{m,\xi} \Lambda^{(\xi,L)}_{n\,m}(\mathbf{k},\mathbf{q}+\mathbf{G})\Lambda^{(\xi,l)}_{m\,n'}(\mathbf{k}+\mathbf{q}, -\mathbf{q}-\mathbf{G}),
	\label{Hbg_F}
\end{align}
where $V_{L l}(\mathbf{q}+\mathbf{G})$ is the interlayer Coulomb interaction at momentum $\mathbf{q}+\mathbf{G}$, and the product of two $\Lambda$ tensors arises from the two exchange vertices: an active electron $n$ at $\mathbf{k}$ scatters into a remote state $m$ at $\mathbf{k}+\mathbf{q}$ (via $\Lambda_{n m}^{(\xi,L)}$ on layer $L$) and then back to an active state $n'$ at $\mathbf{k}$ (via $\Lambda_{m n'}^{(\xi,l)}$ on layer $l$ after exchanging momentum $\mathbf{q}+\mathbf{G}$). The factor $(f_{m,\xi}-\tfrac{1}{2})=\pm\tfrac{1}{2}$ weights the contribution of each remote band $m$ depending on whether it is filled or empty (remote valence or conduction). We emphasize that $\hat{\mathcal{H}}_{\mathrm{bg}}$ is a static one-body Hamiltonian (a sum of layer-dependent potentials and exchange self-energies) that is fixed by the remote-band configuration and does not depend on the active-state occupations $\delta \rho$. In practice this $\hat{\mathcal{H}}_{\mathrm{bg}}$ provides an effective \textbf{background HF potential} (sometimes called the “reference” or “subtraction” potential: that renormalizes the energies of the active bands. 

\subsection{Active terms}
Finally, the term $\hat{\mathcal{H}}_{\mathrm{int}}[\delta \rho]$ represents the interaction corrections from the active subspace itself. It is a functional of the active electron density matrix $\rho_{nn'}$,
We may similarly decompose it into Hartree and Fock pieces: 
\begin{equation}
	[H_{\mathrm{int}}[\delta \rho]]_{n n'}(\mathbf{k}) = [H_{\mathrm{int}}^{H}[\delta \rho]]_{n n'}(\mathbf{k}) + [H_{\mathrm{int}}^{F}[\delta \rho]]_{n n'}(\mathbf{k}),
	\label{Hint_sum}
\end{equation}
where $\delta \rho_{mn} (\mathbf{k},\xi,s) = \rho_{mn} (\mathbf{k},\xi,s) -\frac{1}{2}\mathbb{I}$ is the relative density matrix. Here, $s=\uparrow,\downarrow$ represent spin degree of freedom. This kind representation of density matrix imply we neglect the spin, valley and momentum coherence. For the active Hartree terms, we have
\begin{gather}
	[H_{\mathrm{int}}^{H}[\delta \rho]]_{n n'}(\mathbf{k},\xi,s)= \sum_{\mathbf{G}Ll}V_{Ll}(\mathbf{G})\Lambda^{(\xi,L)}_{nn'}(\mathbf{k},\mathbf{G}) \mathcal{N}_l^\text{act}[\delta\rho]
	\label{ActiveHartree}
\end{gather}
where 
\begin{gather}
	\mathcal{N}_l^\text{act}=\frac{1}{N_\text{cell}}\sum_{k',\xi^\prime,s'}\sum_{m,m'} \Lambda^{(\xi',l)}_{mm'}(\mathbf{k}',-\mathbf{G})\delta\rho_{m'm}(\mathbf{k}',\xi',s')
\end{gather} 
is the self-consistent charge on layer $l$ that is contributed from active bands.
The Fock term is written,
\begin{align}
	&[H_{\mathrm{int}}^{F}[\delta \rho]]_{n n'}(\mathbf{k},\xi,s) = -\sum_{\mathbf{q},\mathbf{G},Ll} \frac{V_{L l}(\mathbf{q}+\mathbf{G})}{N_{\mathrm{cell}}} \label{ActiveFock}\\\nonumber &\times\sum_{\substack{m,m'}}\Lambda^{(\xi,l)}_{m,n}(\mathbf{k}+\mathbf{q}, -\mathbf{q}-\mathbf{G}) \Lambda^{(\xi,L)}_{m',n'}(\mathbf{k},\mathbf{q}+\mathbf{G})\\\nonumber
	&\times\delta\rho_{m'm}(\mathbf{k+q},\xi,s),
\end{align}

In summary, the mean-field Hamiltonian for twisted moiré Bernal graphene (tMBG) is constructed as $\hat{\mathcal{H}}_{\mathrm{HF}} = \hat{\mathcal{H}}_0 + \hat{\mathcal{H}}_{\mathrm{bg}} + \hat{\mathcal{H}}_{\mathrm{int}}[\delta \rho]$, comprising the single-particle term $\hat{\mathcal{H}}_0$, the background HF potential $\hat{\mathcal{H}}_{\mathrm{bg}}$ from remote bands (valence and conduction treated via a half-filling reference as in \cite{kwan2023HF}), and the interaction corrections $\hat{\mathcal{H}}_{\mathrm{int}}[\delta \rho]$ arising from the active bands’ density matrix. The tensors $\Gamma$, $\Lambda$, and $\mathcal{N}_l^{\mathrm{bg}}$ defined in Eqs.~(\ref{Lambda_def})–(\ref{Nbg_def}) encapsulate the moiré form factors and background charge needed to evaluate the Hartree–Fock matrix elements in Eqs.~(\ref{Hbg_H})–\eqref{ActiveFock}. This formalism cleanly separates the static renormalization from remote bands (encoded in $\hat{\mathcal{H}}_{\mathrm{bg}}$) and the self-consistent interactions within the active bands (encoded in $\hat{\mathcal{H}}_{\mathrm{int}}[\delta \rho]$), providing a clear framework for implementing the HF mean-field Hamiltonian of tMBG.

\subsection{Self-consistent Hartree-Fock procedure}

Our self-consistent HF workflow begins with the computation of the form factors associated with each valley. For a given valley, the form factors are classified into two categories based on the band indices: (i) inter-band form factors between the remote bands and the active bands, and (ii) intra-band form factors within the active bands themselves. In our calculations, we designate three conduction bands and three valence bands as the active bands.
Subsequently, we compute the background Hamiltonian. For completeness and numerical stability, we include all bands in the CM model in this stage.
Next, we calculate and storage the Hartree and Fock interaction tensors, denoted respectively as
$
V^H_{\xi, \mathbf{k}, n, n';\, \xi', \mathbf{k}', m', m}
\quad \text{and} \quad
V^F_{\xi, \mathbf{k}, n, n';\, \mathbf{q}, m, m'},
$, which satisfied.
\begin{widetext}
\begin{gather}
	[H_{\mathrm{int}}^{H}[\delta \rho]]_{n n'}(\mathbf{k},\xi,s)= \sum_{\xi', \mathbf{k}', s', m', m} V^H_{\xi, \mathbf{k}, n, n';\, \xi', \mathbf{k}', m', m} \delta\rho_{m'm}(\mathbf{k}',\xi',s')
	\label{eq:H_H}
\\
	[H_{\mathrm{int}}^{F}[\delta \rho]]_{n n'}(\mathbf{k},\xi,s) =-\sum_{\mathbf{q}, m, m'}V^F_{\xi, \mathbf{k}, n, n';\, \mathbf{q}, m, m'} \delta\rho_{m'm}(\mathbf{k+q},\xi,s')
	\label{eq:H_F}
\end{gather}
\end{widetext}
as defined in Eq. \eqref{eq:H_H}-\eqref{eq:H_F}. These tensors serve as the effective interactions in the HF Hamiltonian and are treated as fixed input data during the self-consistency loop.

With the interaction tensors fixed, we iteratively solve for the density matrix $\rho_{m,m'}$ until convergence is achieved, following the standard self-consistent Hartree-Fock procedure.
\begin{gather}
	\rho_{nn'}^{(i)}(\mathbf{k},\xi,s)=\sum_{\alpha\in\text{occ}(\xi,s)} V_{\alpha,n}^{(i)} (\mathbf{k},\xi) V_{{\alpha,n'}}^{(i)\dagger}(\mathbf{k},\xi)
\end{gather}
where $V_{\alpha,n}^{(i)} (\mathbf{k},\xi)$ is the eigenvector of HF Hamiltonian within the ith step iteration, $\alpha$ is sum over the occupated orbitals. In our spin-valley polarized set up at filling $\nu=1$, we separately occupy each $(\xi,s)=(K,\uparrow)\,,(K,\downarrow)\,,(K',\uparrow)\,,(K',\downarrow)$ sectors with $3+1,3+0,3+0,3+0$ active bands for each k-points.
\begin{figure*}
	\centering
	\includegraphics[width=\linewidth]{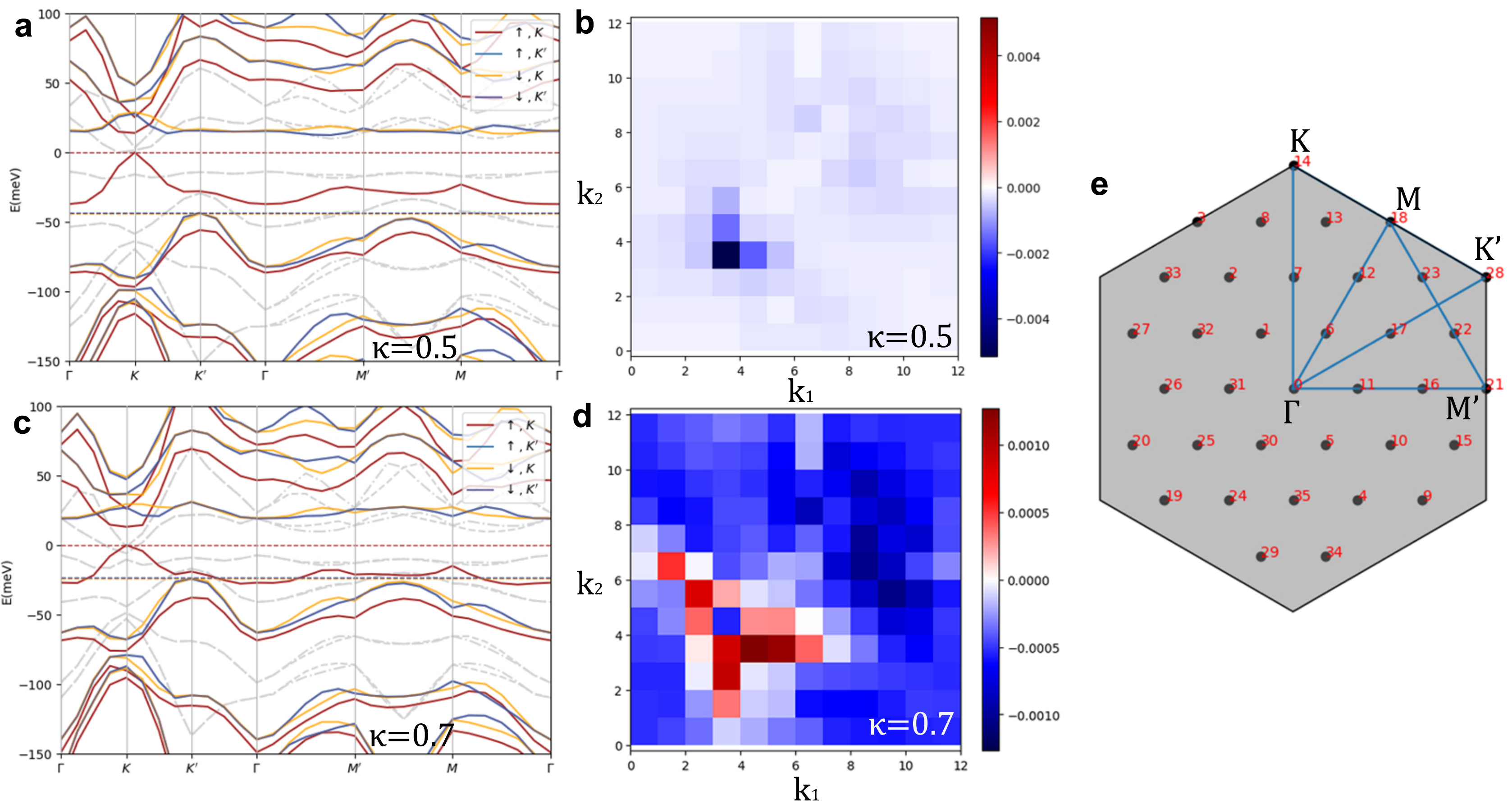}
	\caption{(a,c) Spin-valley polarized Hartree Fock bands on $ N_1\times N_2=12\times12$ mesh,the dashed colored line label the maximum filling of corresponding favor, the gray dashed line represent non-interacting bands. (b, d) Berry curvature distribution of target flat bands, the colorbar labels ${F}/{2\pi N_1 N_2}$. (e) The momentum space mesh for HF calculation, the path along high symmetry points are illustrate, where the path points are labeled with $N_1k_2+k_1$.}
	\label{fig:hfresults}
\end{figure*}
\subsection{Full Hartree Fock band structure}
In this section, we present the fully spin-valley-polarized band structures at $\kappa = 0.5$ and $\kappa = 0.7$. As illustrated in Fig.~\ref{fig:hfresults}(e), these calculations are performed on a $12 \times 12$ momentum mesh within the first Brillouin zone (BZ). For comparison, Figs.~~\ref{fig:hfresults}(b) and (d) display the corresponding Berry curvature distributions. At $\kappa = 0.5$, the Berry curvature of the target conduction bands remains strictly negative throughout the entire BZ, whereas at $\kappa = 0.7$, the Berry curvature exhibits nodes and changes sign, resulting in a net cancellation within the BZ. 
Consistent with the results presented in the main text, the variance of the Berry curvature for the two cases is found to be 0.17 and 0.08, respectively, while the associated Chern numbers of the target conduction bands are $C = 2$ and $C = 1$. The deviations from the trace condition are found to be $\eta=6.8$ and $\eta=9.6$ for $\kappa = 0.5$ and $\kappa = 0.7$, respectively. Compared to the corresponding non-interacting values of $\eta=4.2$ and $\eta=6.3$, this indicates a further departure from the ideal flat-band limit at the Hartree-Fock level. Notably, when accounting for all spin and valley flavors, the interaction clearly favors a single spin-valley polarization near the Fermi level—specifically, the $(K, \uparrow)$ sector—which supports the validity of the single-band projection employed in our ED calculations. Furthermore, since the interactions separate the bands of different spins and valleys, a typical orbital ferromagnetism develops, leading to spontaneous breaking of time-reversal symmetry.

\subsection{Hartree--Fock behavior near the critical region}

For a conventional non-self-consistent single-particle Hamiltonian, a change in the band Chern number under continuous parameter tuning generally requires a band touching at the critical point. In the present Hartree--Fock (HF) calculation, however, the effective single-particle Hamiltonian is obtained self-consistently, so the converged solution does not necessarily evolve along a single continuous branch throughout the transition region.

In practice, we performed repeated scans of $\kappa$ with progressively finer step sizes in the vicinity of $\kappa_c$ on a $12\times 12$ momentum mesh. Within our numerical resolution, we do not resolve an explicit band-gap closing as $\kappa$ approaches the transition. Instead, the numerical results are more naturally interpreted as indicating two competing self-consistent gapped solutions with distinct Chern numbers, $C=2$ and $C=1$, whose energies become very close near $\kappa_c$. As $\kappa$ is tuned across the critical region, the converged HF solution switches between these nearby self-consistent branches, rather than continuously tracing a single band structure through a clearly resolved gap-closing point.

\begin{figure}
	\centering
	\includegraphics[width=\linewidth]{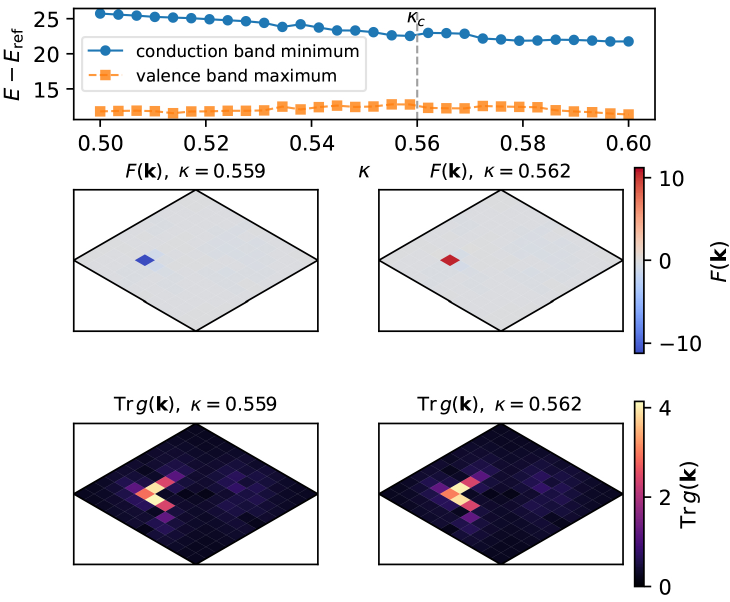}
	\caption{Hartree–Fock results near the critical region. (a,b) Representative HF band structures at $\kappa=0.559$ and $0.562$, on the two sides of the transition. The blue downward and orange upward arrows mark, respectively, the conduction-band minimum and valence-band maximum near $K$, while the black dashed box highlights the target bands. (c) Evolution of the conduction-band minimum and valence-band maximum versus $\kappa$; the dashed vertical lines indicate the critical interval around $\kappa_c$, and repeated fine scans do not resolve an explicit band-gap closing in the converged HF spectra. (d,e) Berry-curvature distribution $F(\mathbf{k})$ for the same two representative $\kappa$ values. (f,g) Corresponding trace of the quantum metric ${\rm Tr}\,g(\mathbf{k})$. The orange arrows in (d–g) indicate the same momentum-space region near $K$ highlighted in (a,b). Across the transition, the Berry curvature near $K$ changes abruptly and reverses sign, while ${\rm Tr}\,g(\mathbf{k})$ is strongly enhanced in the same region.}
	\label{fig:hfkappac}
\end{figure}

For this reason, the most robust HF signature of the transition in our calculation is not a directly observed band touching, but the sharp reconstruction of the band geometry near the $K$ point. As shown in Fig.~\ref{fig:hfkappac}, the Berry curvature of the relevant band changes abruptly across the transition and reverses sign in the vicinity of $K$. At the same time, the trace of the quantum metric is strongly enhanced in the same momentum-space region. These features consistently indicate a pronounced local reconstruction of the Bloch-state geometry near the critical region.

We therefore regard the limiting behavior approached from the two sides of the transition as more informative than a nominally unique HF solution exactly at $\kappa=\kappa_c$. In this sense, the HF calculation is used here as a single-particle diagnostic of geometric reconstruction near the transition, rather than as a direct many-body proof of the FCI transition itself.

\section{Hybrid Wannier Functions}

The hybrid Wannier function (hWF) needs an extra $k_x$–dependent phase to enforce the parallel-transport gauge along $\hat x$ (so that the state is maximally localized in $x$). The corrected single-band expression at fixed $k_y$ is
\begin{align}
	&|W_{nX}(k_y)\rangle=\frac{L_1}{2\pi}\oint_{\mathrm{BZ}_x}\!dk_x\;
	e^{-i k_x X L_1}\nonumber\\
	&\qquad\times\exp\!\left[-\,i\!\int_{0}^{k_x}\!dk_x'\,\big(A_x^{(n)}(k_x',k_y)-\bar A_x^{(n)}(k_y)\big)\right]
	\,|\psi_{n\mathbf k}\rangle,
	\label{eq:hWF-kx-def}
\end{align}
where 
\begin{align}
	A_x^{(n)}(\mathbf k)&=i\langle u_{n\mathbf k}|\partial_{k_x}u_{n\mathbf k}\rangle,\\
	\bar A_x^{(n)}(k_y)&=\frac{1}{2\pi/L_1}\oint_{\mathrm{BZ}_x}\!dk_x\,A_x^{(n)}(\mathbf k)
	=\frac{\phi_n(k_y)}{L_1},\\
	\bar x_n(k_y)&=\frac{L_1}{2\pi}\,\phi_n(k_y),\qquad 
	e^{i\phi_n(k_y)}=\mathrm{eig}\,\mathcal W_x(k_y).
\end{align}
The factor $\exp\!\big[-i\int^{k_x}(A_x-\bar A_x)\,dk_x'\big]$ removes the longitudinal Berry-connection fluctuations along $k_x$ (up to the constant average $\bar A_x$), i.e. imposes the parallel-transport gauge, which is the condition that guarantees maximal localization along $x$.
\subsection{Geometry of Hybrid Wannier function}
In real space we define the hybrid Wannier wave function by
\begin{equation}
	\psi_{nX}(x,y;k_y)
	=\langle x,y|W_{nX}(k_y)\rangle.
\end{equation}
In the plane-wave representation used in the numerics, this can be written schematically as
\begin{equation}
	\psi_{nX}(x,y;k_y)
	=\sum_{G=(g,h)} c^{(n)}_{X,G}(k_y)\,
	e^{i2\pi\big[g\,x + (h+k_y^{\mathrm{frac}})\,y\big]},
	\label{eq:hWF-realspace-expansion}
\end{equation}
where $G=(g,h)$ are integer reciprocal-lattice indices, $k_y^{\mathrm{frac}}=k_y L_2/2\pi$ is the reduced crystal momentum along $\hat y$, and the coefficients $c^{(n)}_{X,G}(k_y)$ are obtained from the Bloch eigenstates $|u_{n\mathbf k}\rangle$ by a discrete Fourier transform in $k_x$ combined with the parallel-transport gauge fixing.

The corresponding probability density in real space is
\begin{equation}
	\rho_{nX}(x,y;k_y)
	=|\psi_{nX}(x,y;k_y)|^2,
\end{equation}
which is normalized as
\begin{equation}
	\int dx\int dy\;\rho_{nX}(x,y;k_y)=1.
\end{equation}
(On a discrete grid $\{x_j,y_i\}$ with spacings $\Delta x,\Delta y$, the normalization condition becomes
$\sum_{i,j}\rho_{nX}(x_j,y_i;k_y)\,\Delta x\,\Delta y=1$.)

\begin{figure*}
	\centering
	\includegraphics[width=\linewidth]{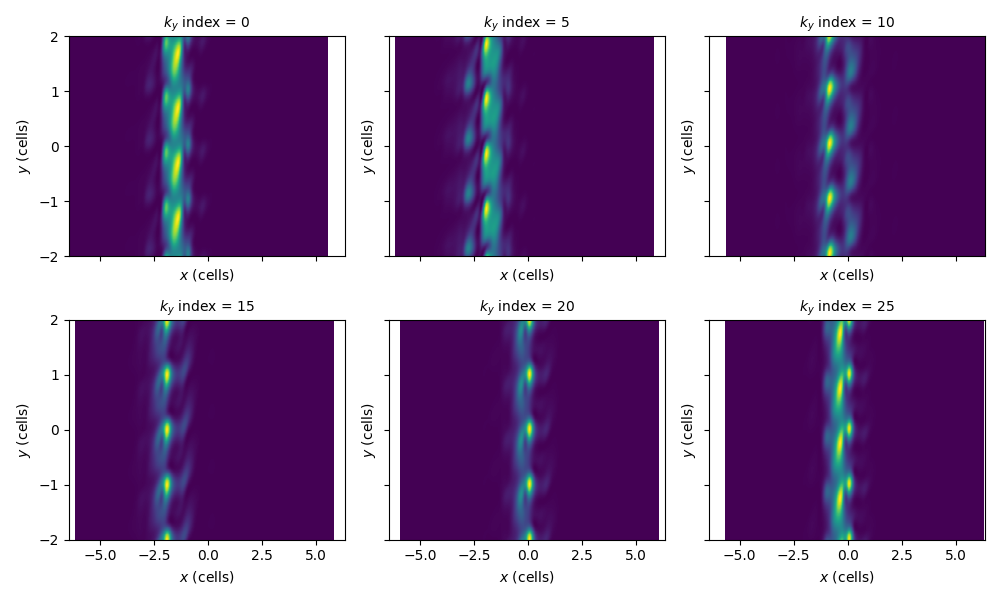}
	\caption{Hybrid Wannier functions distribution for $\kappa=0.64$ in real space at different $k_y$, where the $k_y$ index from $0$ to $30$ indicates the path along $\Gamma-K-M-K'-\Gamma$.}
	\label{fig:hwannierfrealspace}
\end{figure*}

In Figure~\ref{fig:hwannierfrealspace}, we plot the real-space distribution of the hWF of the target band at different $k_y$. In principle, we can define its $x$–direction width using the marginal density
\begin{equation}
	\rho^{(x)}(x;k_y)
	=\int dy\;\rho_{nX}(x,y;k_y),
	\label{eq:marginal-density}
\end{equation}
and the corresponding center
\begin{equation}
	\mu(k_y)
	=\frac{\displaystyle\int dx\;x\,\rho^{(x)}(x;k_y)}
	{\displaystyle\int dx\;\rho^{(x)}(x;k_y)}.
	\label{eq:hWF-center-from-density}
\end{equation}
In the parallel-transport gauge, $\mu(k_y)$ coincides with the Wannier center $\bar x_n(k_y)$ extracted from the Wilson loop, up to the integer cell index $X$.

For this single target band, we denote the $x$–direction standard deviation of the hybrid Wannier function by $w(k_y)$. It is defined as
\begin{equation}
	w^2(k_y)
	=\frac{\displaystyle\int dx\;\big(x-\mu(k_y)\big)^2\,
		\rho^{(x)}(x;k_y)}
	{\displaystyle\int dx\;\rho^{(x)}(x;k_y)},
	\label{eq:hWF-width-continuum}
\end{equation}
and
\begin{equation}
	w(k_y)=\sqrt{w^2(k_y)}.
\end{equation}
On the discrete real-space grid used in the calculation, Eq.~\eqref{eq:hWF-width-continuum} is implemented as
\begin{align}
	\rho^{(x)}(x_j;k_y)
	&=\sum_i \rho_{nX}(x_j,y_i;k_y)\,\Delta y,\\
	\mu(k_y)
	&=\frac{\displaystyle\sum_j x_j\,\rho^{(x)}(x_j;k_y)\,\Delta x}
	{\displaystyle\sum_j \rho^{(x)}(x_j;k_y)\,\Delta x},\\
	w^2(k_y)
	&=\frac{\displaystyle\sum_j \big(x_j-\mu(k_y)\big)^2\,
		\rho^{(x)}(x_j;k_y)\,\Delta x}
	{\displaystyle\sum_j \rho^{(x)}(x_j;k_y)\,\Delta x},
\end{align}
which directly corresponds to the numerical procedure used to extract $w(k_y)$ from the real-space density $\rho_{nX}(x,y;k_y)$.

\subsection{Relation between real-space hWF width and the trace of the quantum metric}

To connect the real-space width $w(k_y)$ defined in Eq.~\eqref{eq:hWF-width-continuum} to the band quantum geometry, we recall that for a single isolated band $n$ the quantum geometric tensor of the cell-periodic Bloch state $|u_{n\mathbf k}\rangle$ is
\begin{equation}
	Q^{(n)}_{\mu\nu}(\mathbf k)
	=\langle\partial_{k_\mu}u_{n\mathbf k}|
	\bigl(1-|u_{n\mathbf k}\rangle\langle u_{n\mathbf k}|\bigr)
	|\partial_{k_\nu}u_{n\mathbf k}\rangle,
\end{equation}
whose real part defines the quantum metric
\begin{equation}
	g^{(n)}_{\mu\nu}(\mathbf k)
	=\mathrm{Re}\,Q^{(n)}_{\mu\nu}(\mathbf k).
\end{equation}
In particular, for the $x$–direction we have
\begin{equation}
	g^{(n)}_{xx}(\mathbf k)
	=\langle\partial_{k_x}u_{n\mathbf k}|
	\partial_{k_x}u_{n\mathbf k}\rangle
	-\bigl|\langle u_{n\mathbf k}|\partial_{k_x}u_{n\mathbf k}\rangle\bigr|^2,
	\label{eq:gxx-def}
\end{equation}
and in two dimensions the trace of the metric is
\begin{equation}
	\mathrm{Tr}\,g^{(n)}(\mathbf k)=g^{(n)}_{xx}(\mathbf k)+g^{(n)}_{yy}(\mathbf k).
\end{equation}

At fixed $k_y$, the width $w^2(k_y)$ defined in Eq.~\eqref{eq:hWF-width-continuum} is simply the variance of the position operator $\hat x$ in the hybrid Wannier state $|W_{nX}(k_y)\rangle$:
\begin{equation}
	w^2(k_y)
	=\langle \hat x^2\rangle_{k_y}-\langle \hat x\rangle_{k_y}^2,
\end{equation}
with expectation values taken in $|W_{nX}(k_y)\rangle$. This form is equivalent to Eq.~\eqref{eq:hWF-width-continuum} once the marginal density~\eqref{eq:marginal-density} and the center~\eqref{eq:hWF-center-from-density} are used.

To relate this variance to the quantum metric, it is convenient to rewrite $|W_{nX}(k_y)\rangle$ in $k_x$–space. From Eq.~\eqref{eq:hWF-kx-def} we introduce the total parallel-transport phase
\begin{equation}
	e^{-i\theta_n(k_x,k_y)}
	:=\exp\!\left[-\,i\!\int_{0}^{k_x}\!dk_x'\,\big(A_x^{(n)}(k_x',k_y)-\bar A_x^{(n)}(k_y)\big)\right],
\end{equation}
and define
\begin{equation}
	|f(k_x;k_y)\rangle
	=e^{-i\theta_n(k_x,k_y)}\,|u_{n\mathbf k}\rangle.
\end{equation}
Using $|\psi_{n\mathbf k}\rangle=e^{i\mathbf k\cdot\hat{\mathbf r}}|u_{n\mathbf k}\rangle$ and focusing on the $x$–direction, Eq.~\eqref{eq:hWF-kx-def} can then be written as
\begin{equation}
	|W_{nX}(k_y)\rangle
	=\frac{L_1}{2\pi}\oint_{\mathrm{BZ}_x} dk_x\;
	e^{-ik_x X L_1}\,e^{ik_x \hat x}
	|f(k_x;k_y)\rangle.
\end{equation}
Using the identity
\begin{equation}
	\hat x\,e^{ik_x\hat x}
	= i\,\partial_{k_x}\bigl(e^{ik_x\hat x}\bigr), 
\end{equation}
and integrating by parts over the periodic Brillouin zone (so that boundary terms vanish), one finds the standard relations (for fixed $k_y$)
\begin{align}
	\langle \hat x\rangle_{k_y}
	&=\langle W_{nX}(k_y)|\hat x|W_{nX}(k_y)\rangle\nonumber\\
	&=\frac{L_1}{2\pi}\oint_{\mathrm{BZ}_x} dk_x\;
	i\,\langle f|\partial_{k_x}f\rangle,
	\label{eq:x-mean-f}\\[2mm]
	\langle \hat x^2\rangle_{k_y}
	&=\langle W_{nX}(k_y)|\hat x^2|W_{nX}(k_y)\rangle\nonumber\\
	&=\frac{L_1}{2\pi}\oint_{\mathrm{BZ}_x} dk_x\;
	\langle\partial_{k_x}f|\partial_{k_x}f\rangle.
	\label{eq:x2-f}
\end{align}
Thus the variance can be written as
\begin{equation}
	w^2(k_y)
	=\frac{L_1}{2\pi}\oint_{\mathrm{BZ}_x} dk_x\left[
	\langle\partial_{k_x}f|\partial_{k_x}f\rangle
	-\bigl|i\langle f|\partial_{k_x}f\rangle\bigr|^2\right].
	\label{eq:w2-f-general}
\end{equation}
Equation~\eqref{eq:w2-f-general} is valid in any gauge. For a \emph{single} isolated band we can now choose a gauge where $\langle u_{n\mathbf k}|\partial_{k_x}u_{n\mathbf k}\rangle$ is purely imaginary, and the parallel-transport phase $\theta_n(k_x,k_y)$ is chosen such that
\begin{equation}
	i\langle f(k_x;k_y)|\partial_{k_x}f(k_x;k_y)\rangle
	=\bar A_x^{(n)}(k_y)
\end{equation}
is a $k_x$–independent constant, equal to the $k_x$–averaged Berry connection
(the same $\bar A_x^{(n)}(k_y)$ that appears in the definition of the hWF).
In this gauge
\begin{equation}
	\partial_{k_x}|f\rangle
	=e^{-i\theta_n}\bigl(\partial_{k_x} 
	- i A_x^{(n)}(\mathbf k) + i\bar A_x^{(n)}(k_y)\bigr)
	|u_{n\mathbf k}\rangle,
\end{equation}
and a short algebra gives
\begin{align}
	&\langle\partial_{k_x}f|\partial_{k_x}f\rangle
	-\bigl|i\langle f|\partial_{k_x}f\rangle\bigr|^2\nonumber\\
	&=\langle\partial_{k_x}u_{n\mathbf k}|
	\partial_{k_x}u_{n\mathbf k}\rangle
	-\bigl|\langle u_{n\mathbf k}|\partial_{k_x}u_{n\mathbf k}\rangle\bigr|^2
	=g^{(n)}_{xx}(\mathbf k),
\end{align}
where we used the definition~\eqref{eq:gxx-def} of the quantum metric component $g^{(n)}_{xx}$. Plugging this back into Eq.~\eqref{eq:w2-f-general}, we obtain the key relation
\begin{equation}
	w^2(k_y)
	=\frac{L_1}{2\pi}\oint_{\mathrm{BZ}_x} dk_x\;
	g^{(n)}_{xx}(k_x,k_y).
	\label{eq:w2-gxx}
\end{equation}
That is, for the hybrid Wannier function constructed in the parallel-transport gauge, the real-space variance along $\hat x$ at fixed $k_y$ is exactly the $k_x$–average of the quantum metric component $g_{xx}$.

If we further average $w^2(k_y)$ over $k_y$, we obtain
\begin{align}
	\overline{w^2}
	&:=\frac{L_2}{2\pi}\oint_{\mathrm{BZ}_y} dk_y\;w^2(k_y)\\
	&=\frac{L_1L_2}{(2\pi)^2}\int_{\mathrm{BZ}} d^2k\;
	g^{(n)}_{xx}(\mathbf k),
	\label{eq:w2-avg-gxx}
\end{align}
while a natural ``trace-metric'' measure of the band geometry is
\begin{align}
	w^2_{\mathrm{tr}g}
	&:=\frac{L_1L_2}{(2\pi)^2}\int_{\mathrm{BZ}} d^2k\;
	\mathrm{Tr}\,g^{(n)}(\mathbf k)\nonumber\\
	&=\frac{L_1L_2}{(2\pi)^2}\int_{\mathrm{BZ}} d^2k\;
	\bigl[g^{(n)}_{xx}(\mathbf k)+g^{(n)}_{yy}(\mathbf k)\bigr].
\end{align}
Since the quantum metric is positive semidefinite, $g^{(n)}_{xx}(\mathbf k)\le\mathrm{Tr}\,g^{(n)}(\mathbf k)$ for all $\mathbf k$, and therefore
\begin{equation}
	\overline{w^2}
	\le w^2_{\mathrm{tr}g},
\end{equation}
with equality when the metric is isotropic, $g^{(n)}_{xx}(\mathbf k)=g^{(n)}_{yy}(\mathbf k)$ everywhere in the Brillouin zone. Thus, in the single-band case the hWF variance $w(k_y)$ provides a real-space realization of the band quantum geometry encoded in the quantum metric, and its Brillouin-zone average is directly controlled by the trace of the quantum metric.

%

\section{Exact Diagonalization results}
\subsection{Momentum space mesh}

For each twist angle we start from the two primitive moiré reciprocal-lattice vectors $\mathbf{b}_{M,1},\mathbf{b}_{M,2}$ ($|\mathbf{b}_{M,1}|=|\mathbf{b}_{M,2}|,\;60^{\circ}$).  A momentum mesh is specified by an integer pair $(N_1,N_2)$—the numbers of grid points sampled along the two primitive directions—and by an $SL(2,\mathbb Z)$ matrix
\begin{gather}
	\tilde M=
	\begin{pmatrix}
		\tilde n_{11} & \tilde n_{12}\\[2pt]
		\tilde n_{21} & \tilde n_{22}
	\end{pmatrix},
	\qquad
	|\det\tilde M|=1 .
\end{gather}
The rows of $\tilde M$ generate a tilted basis
$\mathbf f_{1,2}=\tilde n_{1,2;\,1}\,\mathbf{b}_{M,1}+\tilde n_{1,2;\,2}\,\mathbf{b}_{M,2}$ that spans exactly the same Brillouin zone, but reshapes the rectangular $N_1\times N_2$ mesh into a general parallelogram
Every crystal momentum is then labeled by the integer pair
\begin{gather}
\mathbf k(k_x,k_y)=\tfrac{k_x}{N_1}\mathbf f_1+\tfrac{k_y}{N_2}\mathbf f_2, \nonumber\\
\qquad k_x=0,\dots,N_1-1,\;k_y=0,\dots,N_2-1,
\end{gather}
The parallelogram defined by $\mathbf f_{1,2}$ encloses the minimal translational unit of the mesh.  To characterize its anisotropy we inscribe the John in-ellipse of that parallelogram and record its semi-axes $a\ge b$ and the orientation $\theta$ of the major axis.  The ratio $b/a\le 1$ serves as a dimensionless measure of anisotropy: $b/a\simeq1$ corresponds to an essentially isotropic mesh, whereas smaller $b/a$ signals a more elongated, strongly anisotropic sampling. 

 \begin{figure}
	\centering
	\includegraphics[width=\linewidth]{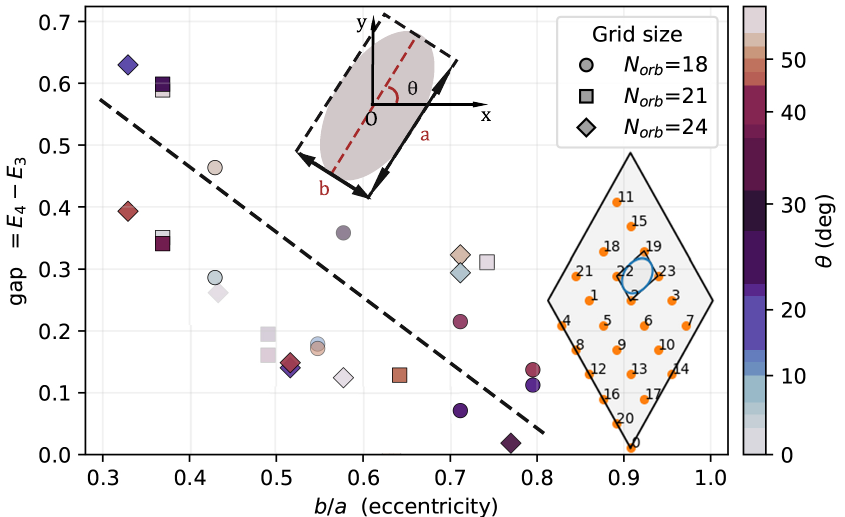} 
	\caption{Variation of ED gap at $\kappa=0.5$ with respect to different momentum-space meshes. Inset: Illustration of a representative mesh at $b/a=0.71,\, \theta=53.1^\circ$. The anisotropic of each mesh is characterized by the eccentricity of the inscribed ellipse within its minimal parallelogram unit.}
	\label{fig:nematicedgap}
\end{figure}

\begin{figure}
	\centering
	\includegraphics[width=\linewidth]{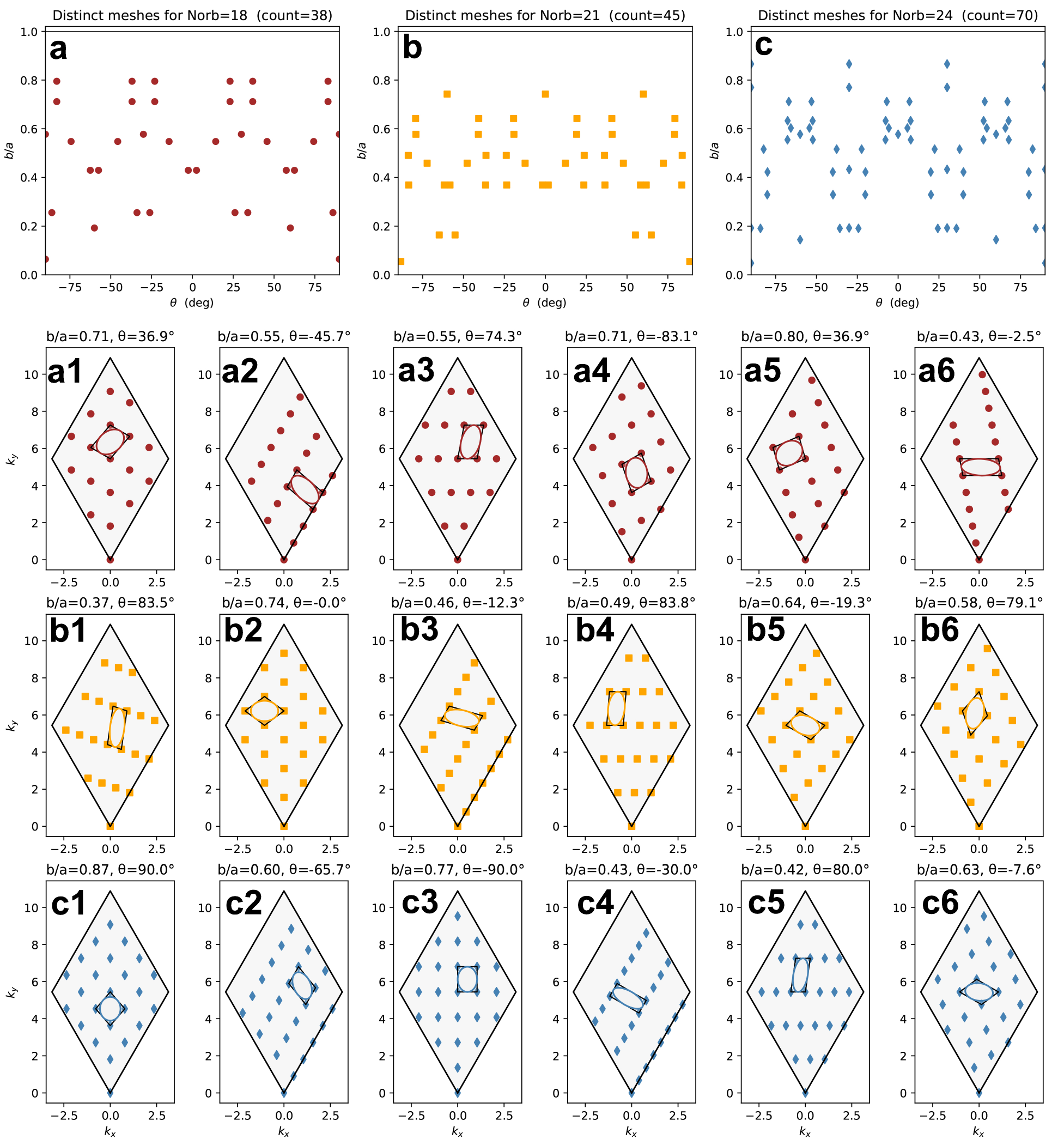}
	\caption{(a, b, c) All Possible mesh at $N_\mathrm{orb}=18,\,21,\,24$. Each points represent a valid mesh. The "count" labels the total number of valid mesh. (ai, bi, ci with $i=1,\cdots, 6$) Examples of different kind of mesh for $N_\mathrm{orb}=18,\,21,\,24$. }
	\label{fig:meshtypes}
\end{figure}

\begin{figure*}
	\centering
	\includegraphics[width=1\linewidth]{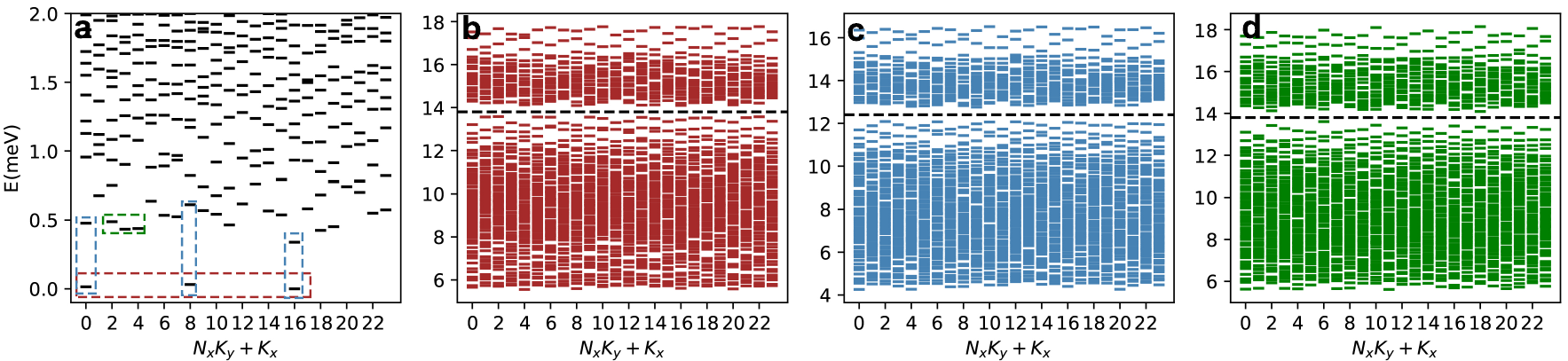}
	\caption{(a) ED spectrum at $\kappa=0.5$, the colored box indicate states that used to generated PES. (b-d) Corresponding PES extracted from different states labeled by colored box in (a), with $N_A=3,N_B=5$ and $D_{(1,2)}(24,3)=1520$ below the gap.}
	\label{fig:pesexstates}
\end{figure*}

	At smaller values of \(\kappa\), the system is in Halperin (112) like-FCI phase. We find the anisotropy of mesh have great influence on the gap of such FCI states, which may brings anisotropy to computed ground states. To be specific, as illustrate in Fig.~\ref{fig:nematicedgap}, FCI gap shrink at more isotropic choice of mesh.
	In particular, we construct a variety of momentum-space meshes, each characterized by a unique parallelogram unit. 

	To quantify the anisotropic of each mesh, we inscribe an ellipse within its minimal parallelogram unit cell and define the eccentricity \(b/a\) of the ellipse as a proxy. A smaller \(b/a\) corresponds to a more elongated unit cell and hence stronger anisotropic character. Remarkably, we find that ED gaps decrease with $b/a$. The gap is defined as the energy difference between the fourth and third lowest many-body levels. Additionally, the orientation of the ellipse $\theta\in[0,\pi/3]$ (i.e., the angle between its major axis and the \(x\)-direction) also influences the FCI gap, though not in a strictly monotonic or causal way. Increasing the sampling density while holding eccentricity fixed (e.g., at \(b/a = 0.71\)) further enhances the stability of the FCI.
	Remarkably, when $b/a$ is small nearly a thin torus \cite{Bernevig2012}, PES shows a cascade gap including CDW, $(1,3)$ and $(1,2)$-counting rule, which can also be regraded as an indirect signature of unobserved $e/3$-quasi-hole excitation origin from the vortex attachment of Halperin (112) typed ground states. 
	
	As illustrate in Fig. \ref{fig:meshtypes} (a-c), we calculate all possible mesh by scanning $\tilde M$ matrix at fixed total number of orbitals $N_\text{orb}$. Each mesh is parameterized by $(b/a,\theta)$. Due to the intrinsic symmetry, the mesh only needs to take $\theta\in(0,60^\circ]$.  Different examples are also shown in \ref{fig:meshtypes} (a1-a6,b1-b6,c1-c6). 
	\subsection{Derivation of the (1,r) admissible counting}
	We consider a one-dimensional ordering of the $N_{\mathrm{orb}}$ orbitals (e.g., Landau-gauge or hybrid-Wannier orbitals on a torus/cylinder) and impose the $(k,r)$ admissible rule: in any set of $r$ consecutive orbitals, there are at most $k$ particles. For $k=1$, this becomes the generalized Pauli principle
	\begin{equation}
		\sum_{\ell=0}^{r-1} n_{j+\ell}\le 1,\qquad \forall j,
		\label{eq:1r_rule}
	\end{equation}
	with $n_j\in\{0,1\}$ and indices understood modulo $N_{\mathrm{orb}}$. This implies that between any two occupied orbitals along the cycle there must be at least $(r-1)$ empty orbitals (zeros). Such admissibility counting underlies the expected PES/quasihole counting for Laughlin-type states and their lattice (FCI) analogues~\cite{Bernevig2012, Regnault2015PES}.
	
	Let $N\equiv N_{\mathrm{orb}}$ and fix the particle number $N_e$. Traverse the ring and label particles in cyclic order $i=1,\dots,N_e$. Define $g_i$ as the number of empty orbitals between particle $i$ and particle $i+1$ (with $N_e+1\equiv 1$). Then
	\begin{equation}
		g_i \ge r-1,\qquad \sum_{i=1}^{N_e} g_i = N - N_e.
		\label{eq:g_constraints}
	\end{equation}
	Introduce nonnegative integers
	\begin{equation}
		x_i \equiv g_i-(r-1)\ge 0.
		\label{eq:x_def}
	\end{equation}
	Summing \eqref{eq:x_def} and using \eqref{eq:g_constraints} gives
	\begin{equation}
		\sum_{i=1}^{N_e} x_i = (N-N_e)-N_e(r-1)=N-rN_e.
		\label{eq:starsbars_eq}
	\end{equation}
	Hence admissible configurations exist only if $N-rN_e\ge 0$.
	
	Now count configurations. First, fix a reference occupied orbital (e.g., declare ``orbital 0 is occupied'' and call it particle $1$). Then the cyclic gap data $\{x_i\}$ uniquely specifies an admissible configuration, because the orbitals are filled by starting at the reference particle and inserting $g_i=x_i+(r-1)$ zeros before placing the next particle, and so on until the cycle closes. Therefore, the number of admissible configurations \emph{with a fixed reference occupied orbital} equals the number of nonnegative integer solutions of \eqref{eq:starsbars_eq}. By stars-and-bars,
	\begin{align}
		\#_{\mathrm{fixed\ ref}}
		&= \binom{(N-rN_e)+N_e-1}{N_e-1} \nonumber\\
		&= \binom{N-(r-1)N_e-1}{N_e-1}.
		\label{eq:fixed_count}
	\end{align}
	
	Finally, restore translational freedom on the ring. There are $N$ choices of the reference orbital position. However, for a given physical configuration, choosing any one of its $N_e$ occupied orbitals as the ``reference occupied orbital'' produces the same configuration in the fixed-reference counting; thus each configuration is counted $N_e$ times. Consequently,
	\begin{equation}
		D_{(1,r)}(N,N_e)
		= \frac{N}{N_e}\,\#_{\mathrm{fixed\ ref}}
		= \frac{N}{N_e}\binom{N-(r-1)N_e-1}{N_e-1}.
		\label{eq:D1r_final}
	\end{equation}
	Restoring $N=N_{\mathrm{orb}}=N_xN_y$, we obtain the equation shown in the main text.
\subsection{Admissible-$(1,2)$ rule in FCI-1}
At first glance, it seems impossible to generate an admissible $(1,2)$ rule in an FCI whose elementary particles are electrons. However, we emphasize that this counting rule can be extracted not only from the ground states, but also from nearby excited states. As illustrated in Fig.~~\ref{fig:pesexstates}, considering only the ground states [Fig.~~\ref{fig:pesexstates}(a,b)] yields the same PES as when excited states are included [Fig.~~\ref{fig:pesexstates}(a,c)], and even when the ground states are entirely excluded [Fig.~~\ref{fig:pesexstates}(a,d)]. This indicates that, in this case, the PES does not truly reflect the $e/3$ quasiparticle excitations of the ground state. The $(1,2)$-admissible counting, should be associated with bosonic neural excitations (i.e. ($e/3,-e/3$) exciton ).

\begin{figure}[b]
	\centering
	\includegraphics[width=\linewidth]{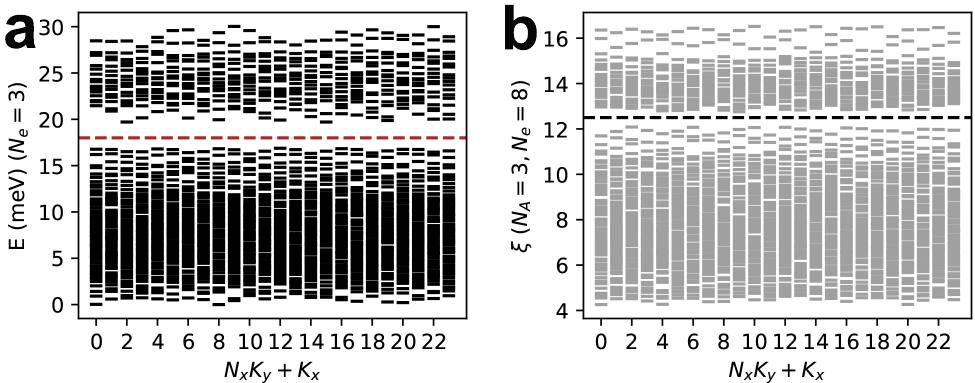}
	\caption{(a) Comparison of ED results with $N_e=3$ on $4\times 6$ regular mesh. (b) PES results at $N_e=8,N_A=3$. All with $1520$ states below the gap, satisfy admissible-(1,2) counting rule.}
	\label{fig:pesvsed}
\end{figure}
\begin{figure*}
	\centering
	\includegraphics[width=0.9\linewidth]{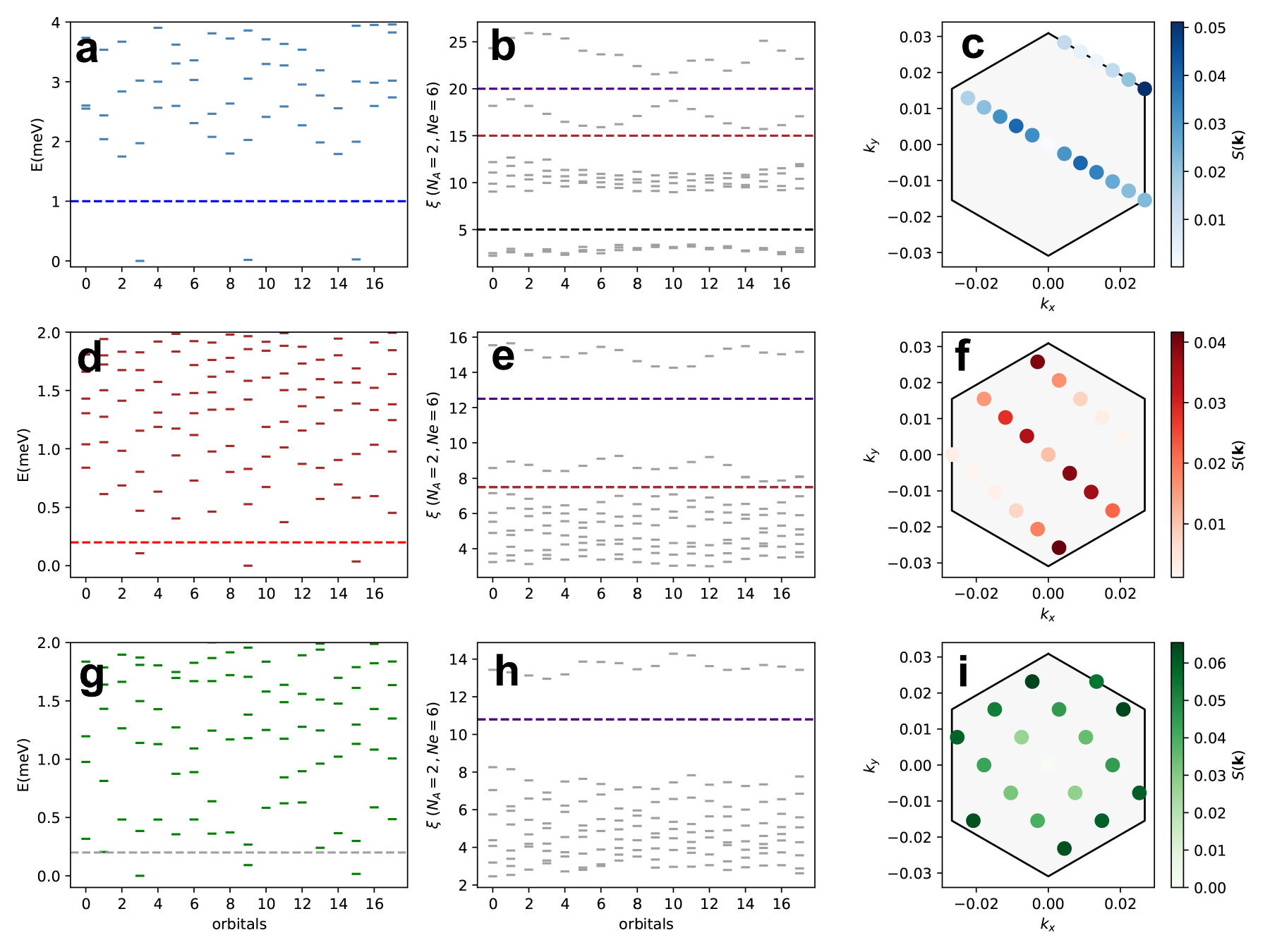}
	\caption{(a-c) for $b/a=0.19$ (d-f) for $b/a=0.41$. (g-h) for $b/a=0.71$ (a, d) $18-$orbitals ED results with 3-folded degenerated ground states in $\nu=1/3$. (b-e) PES with number of levels below black line is $3C_6^3=45$, red line is $117$ as admissible-(1,3) rule  and indigo line is $135$ as (1,2) rule. (c,f,i) The spectral function to detect CDWs. (g) The ED results with $6$ electrons on more isotropic mesh. (f) The PES with only $153$ states below the gap, corresponding to the admissible-(1,2) rule.}
	\label{fig:kappa05fcipessq18}
\end{figure*}

\subsection{Key numerical properties of FCI-1 at $\nu=1/3$}.
For $\kappa<0.555$ at $\nu=1/3$ filling, the system supports FCI that beyond the traditional Lauglin-like FCI. The key feature below,

1. As illustrate in Fig. \ref{fig:kappa05fcipessq18} (a-c), when the mesh holds small eccentricity represent a thin tours i.e. $b/a=0.19$, the PES spectrum supports cascade gaps at $N_A=2$. The first gap indicate a charge density wave (CDW) phase, with the number of states below is $3 C_{N_e=6}^{N_A}$, while the second gap indicate a standard admissible (1,3)-rules, grantee for Lauglin states. Meanwhile, the third gap appears satisfied the admissible (1,2)-rules which is non-trivial for it should indicate bosonic particle-hole excitations. We also consider the spectral function, 
\begin{gather}
S(\mathbf{k})=\frac{1}{N_\text{orb}}(\langle \rho(\mathbf{k})\rho(-\mathbf{k})\rangle-N_e^2\delta_{\mathbf{k},0}	)
\end{gather}
where $\rho(\mathbf{k})=\sum_{\mathbf{k}'} \hat c^\dagger_\mathbf{k+k'}c_\mathbf{k'}$ is density operator. As illustrate in Fig. \ref{fig:kappa05fcipessq18} (c), a peak appear at the $K$ point of the Brillouin zone, consistent with CDW results.

2. As illustrate in Fig. \ref{fig:kappa05fcipessq18} (d-f), when we increase $b/a=0.41$, the CDW gap in PES disappear and the Lauglin-FCI PES gap also shrink. Compared with previous case, the ED gap also decrease. The remaining major gap shows admissible-(1,2) counting. Similarly, in this case, we compute $S(\mathbf{q})$, there is no pronounced peak anymore.

3. As illustrate in Fig. \ref{fig:kappa05fcipessq18} (g-i), when we further make the mesh more isotropic with $b/a=0.80$. we find both the CDW and (1,3)-counting gap closed in PES, with only (1,2)-gap remains. This results indicate the quasi-hole excitations are no-longer gaped with the particle-hole excitations. This can also being checked by performing ED calculation at $N_e=2$ directly. As illustrate in Fig. \ref{fig:pesvsed}, the ED results also gaped and show $153$ states at lower energy sectors, providing a self-consistent prof of admissible (1,2)-counting rule.

\subsection{Different electric field, twist angle, dielectric constant and flux insertion}

As a supplement to the ED phase diagram presented in the main text, we provide additional exact diagonalization results at $\kappa = 0.5$, focusing on specific parameter cuts. 

\begin{figure}
	\centering
	\includegraphics[width=\linewidth]{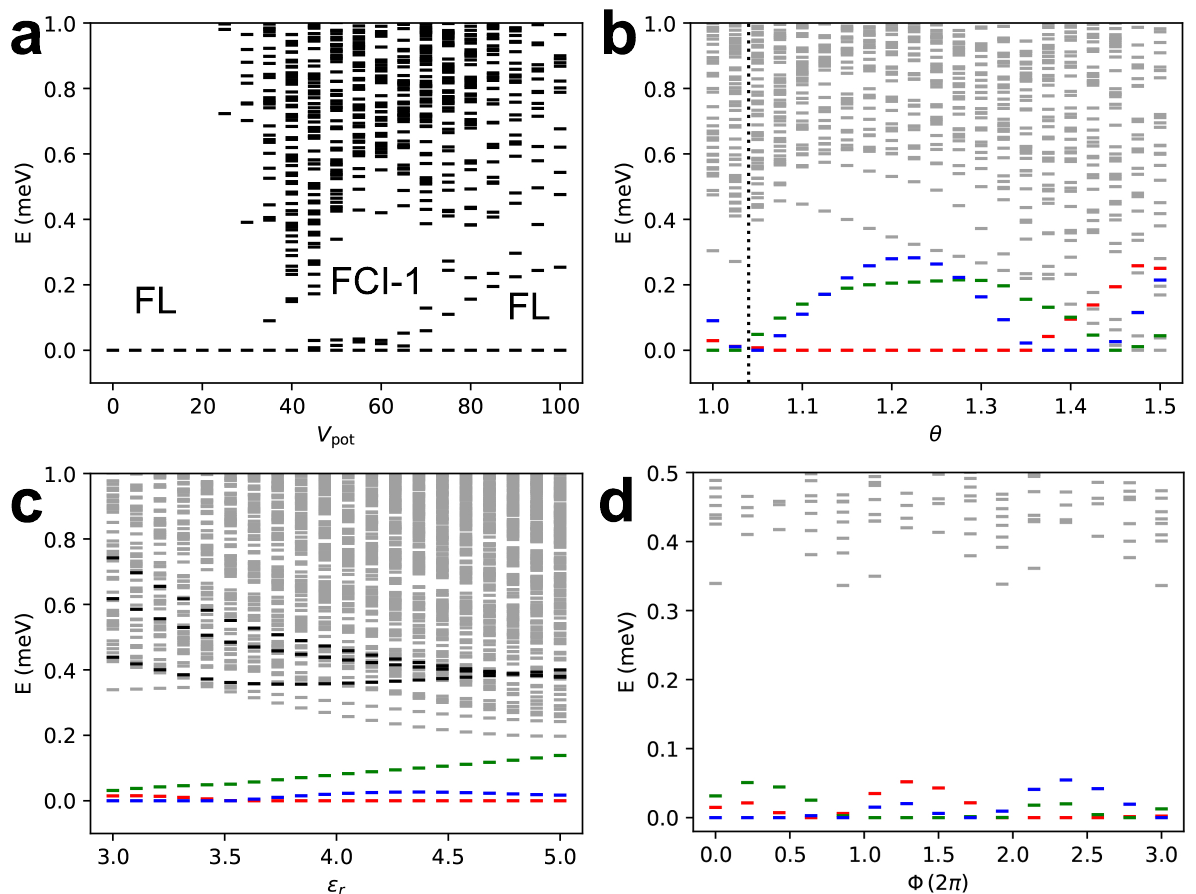}
	\caption{ The variation of ED spectrum 
		(a) in terms of bias;
		(b) in terms of twist angle;
		(c) in terms of dielectric constant.
		(d) The special flow of Harperlin-like FCI.
	}
	\label{fig:edotherparameters}
\end{figure}

In Fig.~\ref{fig:edotherparameters}(a), we show the evolution of the ground state as a function of interlayer potential bias, with all other parameters fixed. A robust FCI ground state is observed only within a narrow bias window between 40–60 meV. Outside this range, the system exhibits a unique ground state, which we identify as a conventional Fermi liquid (FL) phase.

In Fig.~\ref{fig:edotherparameters}(b), we examine the role of the twist angle while keeping the remaining parameters identical to those used in the main text. The FCI-FL-gapless phase transitions are observed by tuning twist angle. The data reveal that the choice of twist angle $\theta = 1.04^\circ$, used in the main text, yields the clearest signature of an FCI: a well-isolated, nearly three-fold degenerate ground state.

Fig.~\ref{fig:edotherparameters}(c) investigates the effect of the dielectric constant. Within the parameter regime used in the main text, we find no qualitative phase transition across the range $\epsilon = 3.0$ to $5.0$. However, the quality of the three-fold degeneracy degrades approximately linearly as the interaction strength weakens, indicating a gradual destabilization of the FCI.

Finally, in Fig.~\ref{fig:edotherparameters}(d), we present the effect of inserting net flux through the system. The stability of the ground state under flux insertion further confirms the robustness of the FCI phase under the parameter choice employed in the main text.

\subsection{Calculation of Many body Chern number}

\begin{figure}
	\centering
	\includegraphics[width=\linewidth]{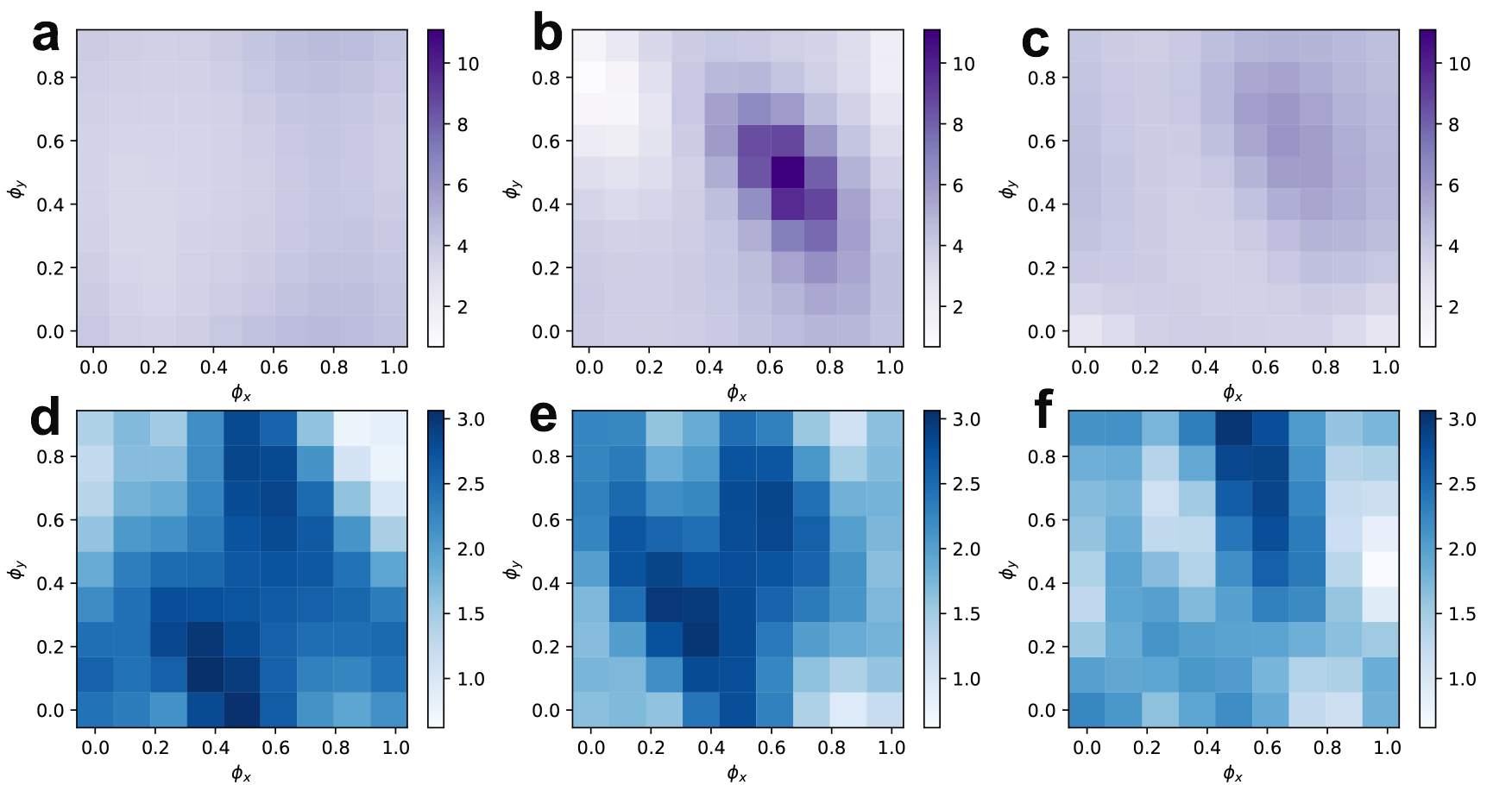}
	\caption{(a-c) The berry curvature for FCI-1 at $\kappa=0.5$ in terms of three ground states, the resulting $C_d\approx2/3$ for each figure is not standard Lauglin-like. Note that (b) shows a pronounced peak for $\kappa=0.5$ is near critical points (d-f) Same results for FCI-2 in $\kappa=0.7$ with $C_d\approx1/3$, which is a standard Lauglin-like results.}
	\label{fig:manybodyberrycurvature}
\end{figure}

Both ED phases are calculated by projecting the interaction onto the original single-particle $C=2$ target band. The threefold ground-state manifolds of FCI-1 at $\kappa<0.555$ and FCI-2 at $\kappa>0.555$ carry total many-body Chern numbers $C_{\mathrm{MB}}\simeq2$ and $C_{\mathrm{MB}}\simeq1$, respectively, and therefore have different Hall conductances. The many-body Chern number is calculated with flux insertion $\mathbf{k+\boldsymbol{\phi}}$ where $\boldsymbol{\phi}=(\phi_x,\phi_y)$ and $\phi_\mu\in[0,2\pi)$. The ground states can therefore be written as,
\begin{gather}
	|\Omega_d(\boldsymbol{\phi})\rangle = \sum_{l} \Omega_{d,l}(\boldsymbol{\phi}) \prod_{\mathbf{k} \in l} \left(\sum_{\alpha,\mathbf{k}} u_{\mathbf{k},\alpha}(\boldsymbol{\phi})\hat c^\dagger_{\mathbf{k},\alpha}|0\rangle \right),
	\label{eq:GS}
\end{gather}
where $d=0,1,2$ is degenerated ground states label, $u_{\mathbf{k},\alpha}(\boldsymbol{\phi})=u_{k_x+\phi_x/(2\pi N_x), k_y +\phi_y/(2\pi N_y),\alpha}$ and $(k_x,k_y)\in(0\cdots N_x, 0\cdots N_y)$ is the single particle Hamiltonian eigenvectors of target flat bands, where $\alpha=(\mathbf{G},\, \sigma,\, L)$ is composite index of reciprocal lattice, sublattices and layers. The core step in calculation is to evaluate the overlap of ground states at different flux insertion, using $\braket{0|\hat c_{\mathbf{k},\alpha} c^\dagger_{\mathbf{k}'\beta}|0}=\delta_\mathbf{k,k}'\delta_{\alpha\beta}$ we have \cite{Okamoto2022},
\begin{align}
	U_d(\boldsymbol{\phi,\phi'}):&=\braket{\Omega_d(\boldsymbol{\phi})|\Omega_d(\boldsymbol{\phi'})} \\\nonumber
	&= \sum_l \Omega^*_{d,l}(\boldsymbol{\phi})\Omega_{d,l}(\boldsymbol{\phi}') 
	\prod_{\mathbf{k}\in l} u^*_{\mathbf{k},\alpha}(\boldsymbol{\phi})u_{\mathbf{k},\alpha}(\boldsymbol{\phi}')
\end{align}
Therefore, the many body Chern number can be computed as,
\begin{align}
	C_{\mathrm{MB}} =\frac{1}{2\pi} \sum_d C_d =\sum_d	\int_{\boldsymbol{\phi}} \mathrm{d}\boldsymbol{\phi} F_d(\boldsymbol{\phi})
\end{align}
with the berry curvature reads,
\begin{align}
	F_d(\boldsymbol{\phi})= &U_d(\boldsymbol{\phi,\phi}+d\phi_x)U_d(\boldsymbol{\phi}+d\phi_x,\boldsymbol{\phi}+d\phi_x+d\phi_y)\nonumber\\
	\times&U_d(\boldsymbol{\phi}+d\phi_x+d\phi_y,\boldsymbol{\phi}+d\phi_y)U_d(\boldsymbol{\phi}+d\phi_y,\boldsymbol{\phi})
\end{align}
where $d\phi_\mu=2\pi/{N_{\phi\mu}}$. Here, we take $N_{\phi x}\times N_{\phi_y} =10\times10$. As illustrate in Fig. \ref{fig:manybodyberrycurvature}, the resulting fractional Chern number $C_d\approx2/3$ for each momentum sector at $\kappa=0.5$ are $C_1=0.62223074$, $C_2=0.69359348$, $C_3=0.68729438$ with total many-body Chern number $C_{\mathrm{MB}}=2.003$ and $C_d\approx1/3$ with $C_1=0.35996735$, $C_2=0.34924193$,  $C_3=0.29389492$ and total many-body Chern number $C_{\mathrm{MB}}=1.003$ for $\kappa=0.7$.

\subsection{Fidelity diagnostic}
\label{app:fidelity}

To diagnose the quantum phase transition we compute the fidelity between
ground states at neighboring values of the tuning parameter $\kappa$.
For each $\kappa$ we diagonalize the many-body Hamiltonian in fixed
total-momentum sectors $K$, and denote by
\begin{equation}
	\ket{\Psi_K(\kappa)}
	= \sum_{\{n\}} C_K(\{n\};\kappa)\,\ket{\{n\}}
\end{equation}
the lowest-energy state in sector $K$, expanded in the occupation-number
basis $\{\ket{\{n\}}\}$ of the ED Hilbert space. In our problem the
topological ground state is three-fold degenerate, with one state in
each of the momentum sectors $K \in K_D$ (typically $|K_D|=3$).

Because of this exact (or near) degeneracy the individual vectors
$\ket{\Psi_K(\kappa)}$ are only defined up to a $\kappa$-dependent
$U(|K_D|)$ rotation within the ground-state manifold. To obtain a
gauge-invariant quantity we work with the equal-weight mixed density
matrix
\begin{equation}
	\rho(\kappa)
	= \frac{1}{N_{\mathrm{GS}}}
	\sum_{K\in K_D} 
	\ket{\Psi_K(\kappa)}\bra{\Psi_K(\kappa)}\,,
	\qquad N_{\mathrm{GS}} = |K_D| \, .
\end{equation}
The fidelity between two nearby points $\kappa$ and
$\kappa+\delta\kappa$ is defined as the Uhlmann fidelity between the
corresponding density matrices,
\begin{equation}
	F(\kappa,\kappa+\delta\kappa)
	= \left[
	\mathrm{Tr}\,
	\sqrt{ \sqrt{\rho(\kappa)}\,
		\rho(\kappa+\delta\kappa)\,
		\sqrt{\rho(\kappa)} }
	\right]^2 .
	\label{eq:uhlmann_fidelity}
\end{equation}
In our case states belonging to different total-momentum sectors are
orthogonal, so $\rho(\kappa)$ is a direct sum of rank-one projectors.
Equation~\eqref{eq:uhlmann_fidelity} then simplifies to
\begin{equation}
	F(\kappa,\kappa+\delta\kappa)
	= \frac{1}{N_{\mathrm{GS}}^2}
	\left(
	\sum_{K\in K_D}
	\bigl|
	{\Psi_K(\kappa)}\rangle\langle{\Psi_K(\kappa+\delta\kappa)}
	\bigr|
	\right)^2 ,
	\label{eq:fidelity_simplified}
\end{equation}
so that the only input needed is the overlap between the many-body
ground states in the same momentum sector.

These overlaps are evaluated efficiently by expressing the ground states
in terms of the $\kappa$-dependent single-particle eigenvectors.
Denoting by $u_{k a}(\kappa)$ the single-particle Bloch eigenvectors
(labelled by orbital index $k$ and band index $a$), we first compute the
diagonal single-particle overlaps
\begin{equation}
	s_k(\kappa)
	= \sum_a u_{k a}^*(\kappa)\,u_{k a}(\kappa+\delta\kappa) \, .
\end{equation}
For a many-body basis state
$\ket{\{n\}} = \prod_k (c_k^\dagger)^{n_k}\ket{0}$ with occupations
$n_k \in \{0,1\}$, the corresponding diagonal overlap factor is
\begin{equation}
	S_{\{n\}}(\kappa)
	= \prod_{k\,|\,n_k=1} s_k(\kappa) \, .
\end{equation}
Inserting this into the expansion of $\ket{\Psi_K(\kappa)}$ and
$\ket{\Psi_K(\kappa+\delta\kappa)}$ in the occupation basis, we obtain

\begin{align}
		&\braket{\Psi_K(\kappa){\Psi_K(\kappa+\delta\kappa)}}\nonumber\\
		&= \sum_{\{n\}}
		C_K^*(\{n\};\kappa)\,
		S_{\{n\}}(\kappa)\,
		C_K(\{n\};\kappa+\delta\kappa) \, .
\end{align}

Finally, these sector-resolved overlaps are combined according to
Eq.~\eqref{eq:fidelity_simplified} to produce the density-matrix
fidelity $F(\kappa,\kappa+\delta\kappa)$ used in the main text.
\subsection{Level crossing near critical points.}
\begin{figure}
	\centering
	\includegraphics[width=\linewidth]{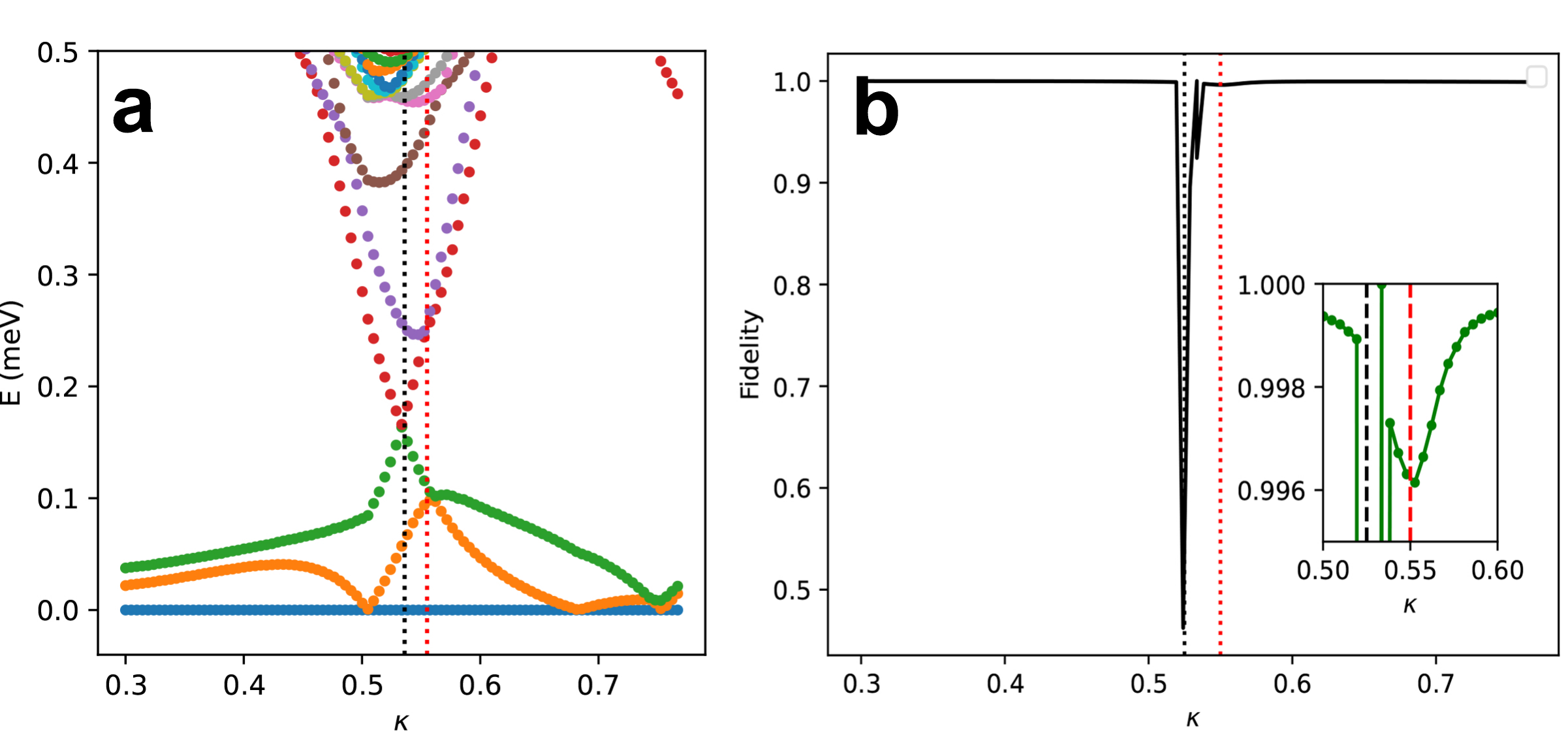}
	\caption{(a) Transition of FCI under flux insertion $\phi=(0.525,0.685)$ in the unit of $2\pi$. Other parameters are consist with Fig. 2 in the main text. (b) Density fidelity under such flux insertion, the black dashed line denotes the discontinuous change of fidelity, and the red dashed line show a dip in inset figure.  }
	\label{fig:largecritical}
\end{figure}
Near a critical point, characterizing topological phase transitions in finite-size systems is often intricate. While the fidelity of the system exhibits typical features of a continuous second-order transition in the absence of flux insertion shown in the main text, the situation changes when a flux is applied. Since a change in the Chern number necessitates the closure of the energy gap under flux insertion, the system manifests characteristics of a first-order transition at specific flux values where the gap closes—here identified numerically at $\Phi = (0.525, 0.685)$. As shown in Fig. \ref{fig:largecritical} (a), the transition point associated with this first-order behavior ($\kappa = 0.536$) slightly deviates from that of the observed second-order transition ($\kappa = 0.555$).  This is further corroborated by the fidelity behavior in Fig. \ref{fig:largecritical} (b): at $\kappa = 0.536$, a sharp jump occurs under the specific flux, whereas at $\kappa = 0.555$, a dip emerges.
\subsection{Calculation details on Particle Entanglement Spectrum}
To determine the topological phase of the ground states in our FCI system, we utilized the particle entanglement spectrum (PES) as a diagnostic tool, recognized for its ability to fingerprint topological order and indicate quasiparticle excitations. The ground states were expressed as \(\lvert \Omega_d \rangle = \sum_i e^{-\xi_i / 2} \lvert \Psi_i^A \rangle \otimes \lvert \Psi_i^B \rangle\), where \(e^{-\xi_i}\) and \(\lvert \Psi_i^A \rangle\) are the eigenvalues and eigenstates of the reduced density matrix \(\rho_A = \mathrm{Tr}_B(\rho)\), with the total density matrix defined by \(\rho = \frac{1}{3} \sum_d \lvert \Omega_d \rangle \langle \Omega_d \rvert\). 

The states \(\lvert \Psi_i^A \rangle\) were constructed block-diagonally within the bases \(\lvert l_A \rangle = \prod_{\mathbf{k} \in l_A} \eta_{k}^\dagger \lvert 0 \rangle\), representing configurations of \(N_A\) particles occupying orbital labels \(\mathbf{k}\) with total momentum \(\mathbf{K}_A = \sum_{\mathbf{k} \in l_A} \mathbf{k}\). For each ground state sector with total momentum \(\mathbf{K}_d\), the complementary momentum \(\mathbf{K}_B = \mathbf{K}_d - \mathbf{K}_A\) defined the bases \(\lvert l_B \rangle\) for \(N_B = N_e - N_A\) particles. In fermionic systems, the combined bases states were given by \(\lvert l \rangle = (-1)^{s_{AB}} \lvert l_A \rangle \otimes \lvert l_B \rangle\), where \(s_{AB}\) accounts for the fermionic sign from the ordering of orbital labels \(\{k_1^A, \ldots, k_{N_A}^A, k_1^B, \ldots, k_{N_B}^B\}\), with each orbital label \(k = N_x k_y + k_x\) enforcing the normal ordering \(\eta_{k_1}^\dagger \cdots \eta_{k_{N_e}}^\dagger \eta_{k_{N_e}} \cdots \eta_{k_1}\) for \(k_1 < k_2 < \ldots < k_{N_e}\). The reduced density matrix elements were calculated as \([\rho_{A}]_{ij} = \sum_{l_B} (-1)^{(s_{A_iB}+s_{A_jB})} \Omega_{\{l_{A_i}\otimes l_B\}}^* \Omega_{\{l_{A_j}\otimes l_B\}}\), where \(\lvert \Omega \rangle = \sum_l \Omega_l \lvert l \rangle\).

PES was computed independently across momentum sectors \(\mathbf{K}_A\), with \(K_{Ax}\) ranging from 0 to \(N_x\) and \(K_{Ay}\) from 0 to \(N_y\). The spectra were analyzed using $(k=1,r=2)$-admissible partition rules, which restrict configurations to no more than one fermion per orbital and no more than $k$ fermions in $k+r$ consecutive orbitals. bases states \(\lvert l_v \rangle\) satisfying these criteria for \(N_A\) particles within an \(N_x \times N_y\) orbital mesh were enumerated. A PES exhibiting an entanglement gap, where the number of entanglement energies below the gap matches the number of valid bases states \(\lvert l_v \rangle\), provided strong evidence of the ground states residing in the FCI phase.

\begin{figure*}
	\centering
	\includegraphics[width=1\linewidth]{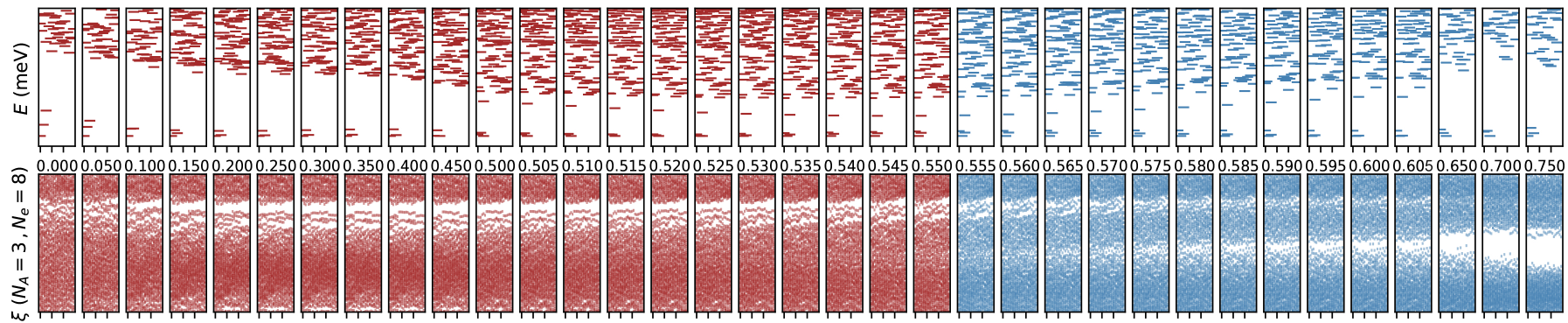}
	\caption{Details on ED (Upper) and PES (lower) transition in the main text Fig. 2,  where x-axis is total momentum label $K_x+N_xK_y$ with $(N_x,N_y)=(4\times6)$. The number between two rows are $\kappa\in[0,0.75]$. Red and blue colors represent FCI-1 and FCI-2. }
	\label{fig:kappavarypes}
\end{figure*}
Here, we present detailed results on the evolution of the particle entanglement spectrum (PES) in Fig. \ref{fig:kappavarypes}. Specifically, for $\kappa$ ranging from $0$ to approximately $0.55$, we observe a well-defined entanglement gap near $\xi \approx 13.8$, below which there are exactly $1520$ states. This counting is consistent with the (1,2)-generalized Pauli exclusion principle, which is not typical-Lauglin type in $\nu=1/3$ for fermions.
As $\kappa$ increases, the trace condition deteriorates further, suggesting the eventual breakdown of the FCI phase. However, intriguingly, the PES shows the emergence of  entanglement gap around $\xi \approx 11$. Below this gap, there are exactly 1088 levels, matching the generalized Pauli principle where no more than one particle occupies any triplet of consecutive orbitals. This level counting aligns precisely with that expected from a Laughlin-type state, indicating a transition in the underlying exclusion statistics.

\section{Landau level expansion under magnetic field}
In the presence of a perpendicular magnetic field, the band structure of tMBG develops Landau quantization and fractal minibands characteristic of the Hofstadter spectrum. To describe this, we define the guiding center coordinate along $y$ as
\begin{equation}
	Y_{M,j} = k_x l^2 + (M p + j) \Delta,
	\label{eq:YM}
\end{equation}
where $l = \sqrt{\hbar / eB}$ is the magnetic length, $p$ and $q$ are integers specifying the magnetic flux per moir\'e unit cell $\Phi/\Phi_0 = p/6q$, and $\Delta=Q_x l^2$ is the $y$-direction spacing between adjacent guiding centers \cite{Bistritzer2011Butterfly}.

The intralayer Hamiltonian in the Landau level basis reads
\begin{equation}
	h_\theta^\pm = -\hbar \omega_c e^{\pm i\theta/2}
	\begin{bmatrix}
			0 & \delta_{n+1,n} \\
			\delta_{n,n+1} & 0
		\end{bmatrix} \otimes \sum_Y |Y L\rangle \langle Y L|,
	\label{eq:hIntra}
\end{equation}
with $\omega_c = \sqrt{2} v_0 / l$ the cyclotron frequency, and $n$ the Landau level index.

The AB-stacked interlayer coupling becomes
\begin{align}
	G &=
	\begin{bmatrix}
			-\hbar v_4 \sqrt{2}/l \delta_{n,n+1} e^{-i\theta/2} & \gamma_1 \\
			-\hbar v_3 \sqrt{2}/l \delta_{n+1,n} e^{-i\theta/2} & -\hbar v_4 \sqrt{2}/l \delta_{n,n+1} e^{-i\theta/2}
		\end{bmatrix} \nonumber
	\\
	&\otimes \sum_Y |Y2\rangle \langle Y1| + \text{h.c.},
	\label{eq:G_LL}
\end{align}
where $|Y1\rangle$, $|Y2\rangle$ denote states on the first and second layers.

The moir\'e interlayer coupling becomes momentum dependent and takes the form
\begin{equation}
	T(k_x, k_y) = \sum_{m,n,\alpha,\beta,j} \sum_{s=1}^3 T_j^{(s)} |2m\alpha j\rangle \langle 1n\beta j|,
\end{equation}
with $T_j^{(s)}$ the tunneling matrices associated with momentum transfer $\mathbf{q}_s$, and matrix elements given by
\begin{align}
	T_j^{(1)} &= T_1 F_{mn}(z_1) e^{-ik_\theta y_0} e^{-4\pi i j q/p}, \\
	T_j^{(2)} &= T_2 F_{mn}(z_2) e^{i k_y \Delta} e^{i k_\theta y_0} e^{i\pi (2j - 1) q/p}, \\
	T_j^{(3)} &= T_3 F_{mn}(z_3) e^{-i k_y \Delta} e^{i k_\theta y_0} e^{i\pi (2j + 1) q/p},
\end{align}
where $y_0 = k_x l^2$ and $z_s = \frac{l}{\sqrt{2}}(q_{s,x} + i q_{s,y})$. The function $F_{mn}(z)$ is the generalized Landau level form factor
\begin{equation}
	F_{mn}(z) = \langle m | e^{z^* \hat{a}^\dagger - z \hat{a}} | n \rangle,
\end{equation}
with closed-form expressions involving associated Laguerre polynomials:
\begin{equation}
	F_{mn}(z) =
	\begin{cases} 
			\sqrt{\frac{n!}{m!}} (z^*)^{m-n} L_n^{(m-n)}(|z|^2) e^{-\frac{|z|^2}{2}}, & m \ge n, \\
			\sqrt{\frac{m!}{n!}} (-z)^{n-m} L_m^{(n-m)}(|z|^2) e^{-\frac{|z|^2}{2}}, & n > m.
		\end{cases}
\end{equation}

The full Hamiltonian has dimension $6Np$, where $N$ is the number of Landau levels kept, and $6 = 2$ (sublattices) $\times$ 3 (layers). For the $-K$ valley, one can obtain the corresponding Hamiltonian by flipping the signs of magnetic flux.

\begin{figure*}
	\centering
	\includegraphics[width=1\linewidth]{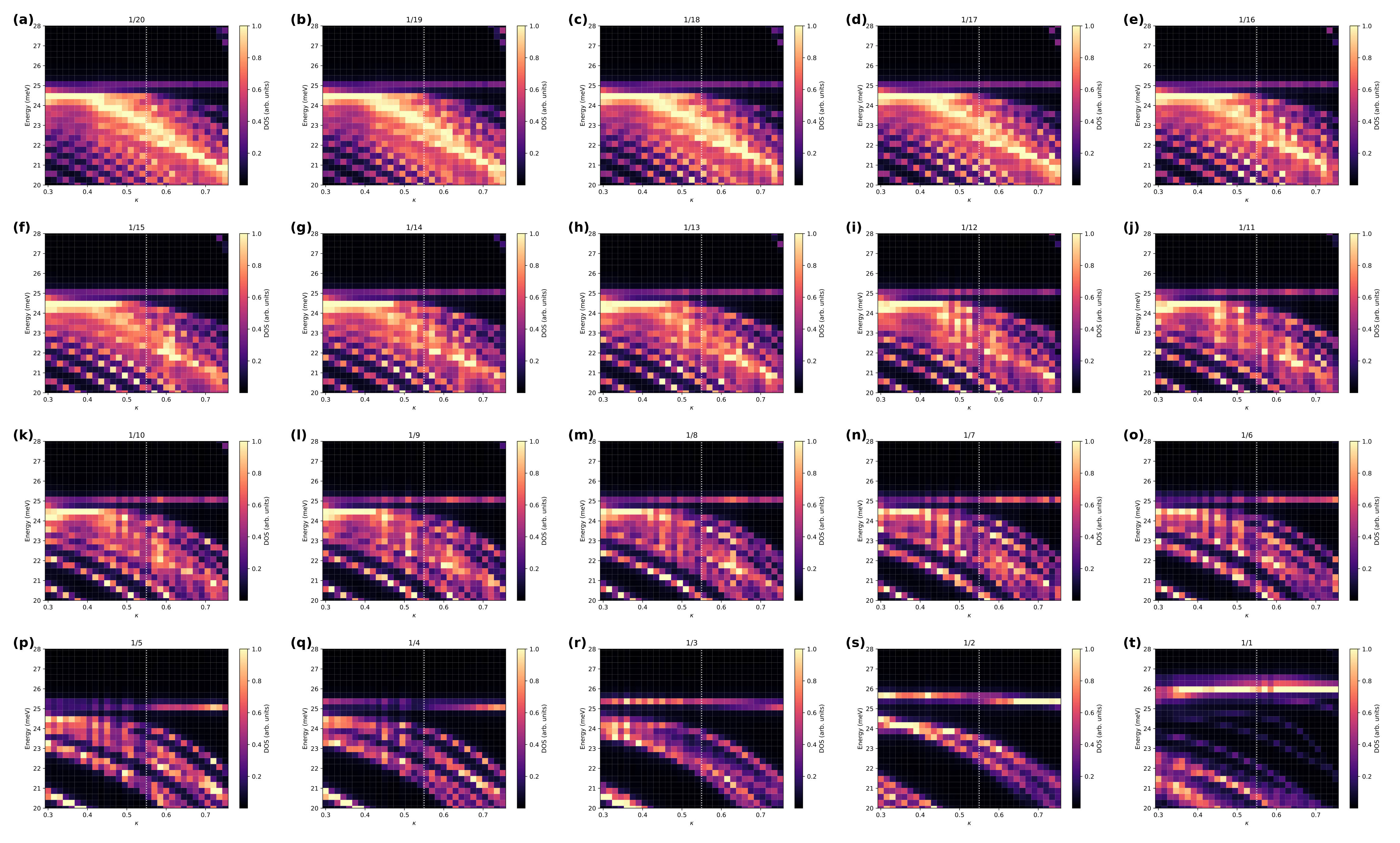}
	\caption{(a-t) The variation of density of states in terms of $\kappa$ for different magnetic flux from $\phi/\phi_0=1/20,1/19,...,1/2,1/1$}
	\label{fig:combinedq20toq1labeled}
\end{figure*}

\subsection{Quantum geometry under magnetic field}
At finite magnetic field the computation of band geometry differs in two essential ways from the zero-field case. First, the presence of magnetic translations implies that Bloch’s theorem holds only in a magnetic Brillouin zone (MBZ) defined at rational flux $\phi=p/q$, with twisted boundary conditions for the periodic part of the wavefunction. Concretely, the commuting set of magnetic translations emerges only after enlarging the real-space unit cell by a factor of $q$, and eigenstates $|\psi_{a\mathbf{k}}\rangle$ then obey $M(\mathbf{R})|\psi_{a\mathbf{k}}\rangle=e^{i\mathbf{k}\cdot\mathbf{R}}|\psi_{a\mathbf{k}}\rangle$ with $\mathbf{k}$ restricted to the MBZ (there are $q$ times as many bands at each $\mathbf{k}$). This MBZ folding and the associated twisted phases must be respected when discretizing $k$-space for geometry.

Second—and crucial for practical evaluation—the use of a Landau-level basis injects \emph{intrinsic curvature} from the basis itself, so the standard zero-field overlap formulas cannot be used naively. As shown in the supplemental derivation, the Landau-level basis carries an analytic, band-diagonal Berry-curvature contribution; hence both the Berry curvature and the quantum geometric tensor (QGT) require a modified, gauge-covariant construction. In the single-band (U(1)) limit this appears as an explicit correction $F=F_{\mathrm{numerical}}-1/p$, but for the general multi-band (U($N$)) case the answer is not a simple “numerical plus analytic” sum and must be extracted from a Wilson-loop formulation \cite{Xie2021TBGFCI}. 

A numerically stable and gauge-invariant route is to compute the non-Abelian Berry curvature from the holonomy of a minimal Wilson loop on the $k$-mesh,
$e^{\,iF(\mathbf{k})\,\Delta k_x\Delta k_y}=W_\square(\mathbf{k})
=U_{\mathbf{k},\mathbf{k}+\Delta_x}\,U_{\mathbf{k}+\Delta_x,\mathbf{k}+\Delta_x+\Delta_y}\,
U_{\mathbf{k}+\Delta_x+\Delta_y,\mathbf{k}+\Delta_y}\,U_{\mathbf{k}+\Delta_y,\mathbf{k}},$
with link matrices $[U_{\mathbf{k},\mathbf{k}'}]_{ab}=\langle U_a(\mathbf{k})|U_b(\mathbf{k}')\rangle$ built from an orthonormal frame $\{|U_a(\mathbf{k})\rangle\}$ spanning the target band subspace. In a Landau-level implementation the eigenstates are expanded as $|U_a(\mathbf{k})\rangle=\sum_I \psi^I_a(\mathbf{k})\,|u_I(\mathbf{k})\rangle$ where $I$ collects internal labels (layer/sublattice/spin), the Landau-level index $n$, and the guiding-center index $j$. The corresponding link overlaps factor into a \emph{numerical} part from the coefficients $\psi^I_a$ and an \emph{analytic} part from the Landau-level transport, so that one must separate the LL analytic parallel transport on each link before forming the plaquette \cite{Xie2021TBGFCI}.

\begin{figure*}
	\centering
	\includegraphics[width=0.8\linewidth]{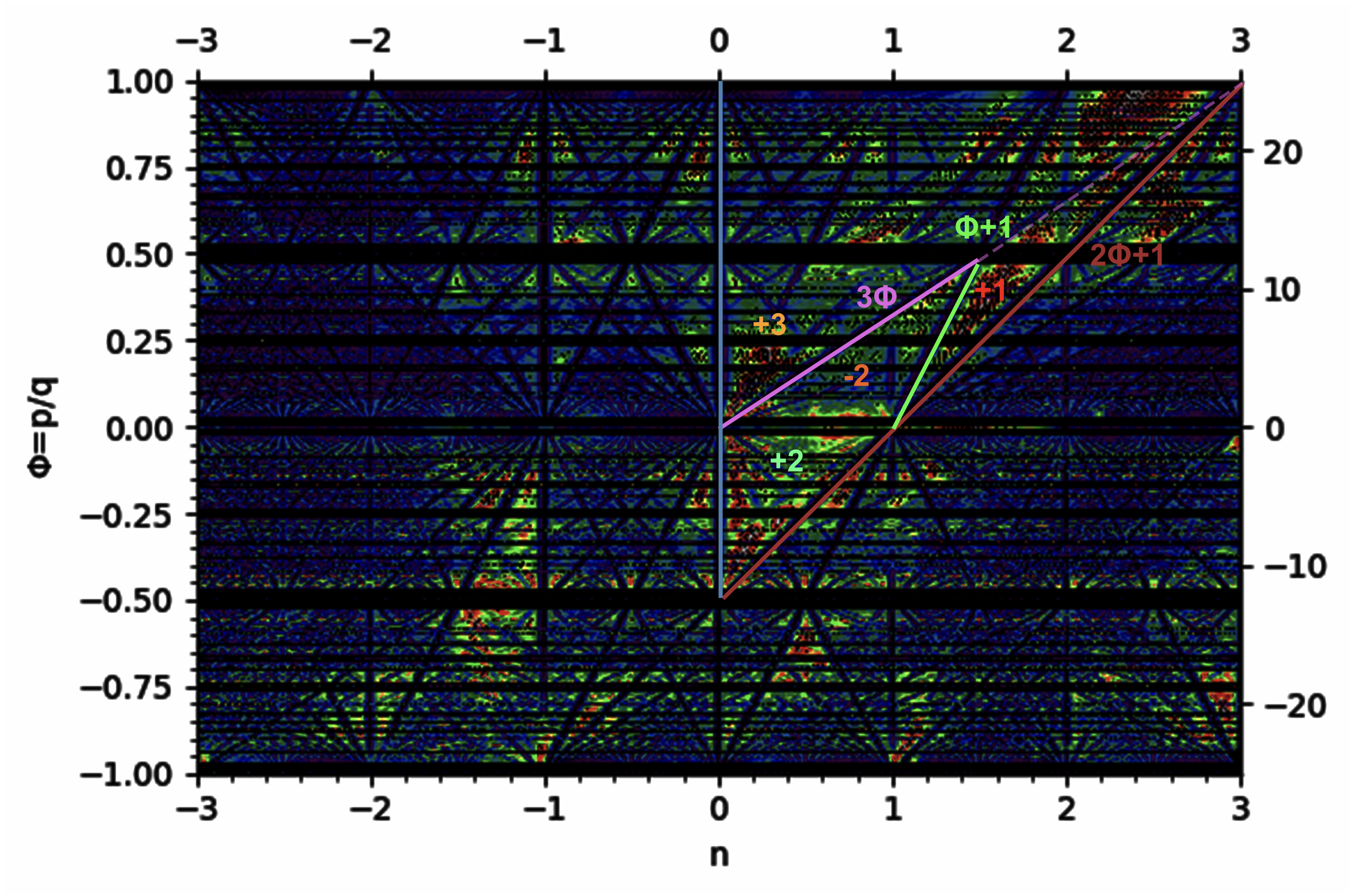}
	\caption{The Wannier plot of Hofstadter butterfly spectrum. The colored line gives $n=C\Phi+\nu$ represent Diophantine equations of gaps. The Chern number of states are also labeled. The highlighted area is mainly considered region in the main text.}
	\label{fig:wannierplot}
\end{figure*}

Operationally, this factorization appears explicitly in the $x/y$-link matrices. In Landau gauge, the $y$-link carries an extra exponential phase that is precisely the band-diagonal LL correction; upon taking the plaquette and tracing the holonomy, this yields a net shift of $(2\pi)^{-1}\!\int \mathrm{tr}\,F$ by $1/p$ per band in the U(1) limit. The full non-Abelian Wilson operator for a plaquette on an $N_x\times N_y$ mesh at flux $p/q$ can be written
\begin{widetext}
	\begin{gather}
		W_\square
		= e^{-i\,\frac{2\pi}{N_xN_y\,p}}\,
		(V_x)_{\mathbf{k},\mathbf{k}+\Delta_x}\,
		(V_y)_{\mathbf{k}+\Delta_x,\mathbf{k}+\Delta_x+\Delta_y}\,
		(V_x^\dagger)_{\mathbf{k}+\Delta_x+\Delta_y,\mathbf{k}+\Delta_y}\,
		(V_y^\dagger)_{\mathbf{k}+\Delta_y,\mathbf{k}},
	\end{gather}
\end{widetext}
with $(V_\alpha)_{\mathbf{k},\mathbf{k}'}=\sum_{I,J}\psi^I_a(\mathbf{k})\,T_\alpha^{IJ}\,\psi^{J}_b(\mathbf{k}')$ and $T_\alpha$ the analytic LL transport matrices on the corresponding link \cite{Xie2021TBGFCI}. This expression implements the necessary subtraction/compensation of the LL-basis curvature in a gauge-covariant way and provides a robust numerical formula for the Berry curvature (from the eigenphases of $W_\square$) on the MBZ. 
The quantum metric under field is obtained analogously by evaluating suitable “back-and-forth’’ Wilson paths (e.g. $k\!\to\!k+q_\alpha e_\alpha\!\to\!k$) with the same LL analytic factors included on each link; in this way both curvature and metric are computed consistently from the same non-Abelian transport that cleanly separates the LL contribution from the numerically determined band mixing. 

Finally, we note practical considerations specific to the LL expansion: the single-particle Hamiltonian at fixed $\mathbf{k}$ is countably infinite-dimensional in the LL index and must be truncated at $n,m<N$; the resulting matrix elements inherit Laguerre-form-factor structure that decays rapidly with $|m-n|$, ensuring numerical tractability while preserving the analytic LL transport used above. These implementation details are orthogonal to the geometry formulas but crucial for stable evaluation of $W_\square$ in large-$q$ MBZs.

\subsection{Landau fan of Wannier plot}
We construct a Wannier-style Landau fan by scanning rational fluxes $\phi=p/q$ and, at each $\phi$, computing the single-particle spectrum $\{\varepsilon_{n}(\mathbf{k};\phi)\}$ over the MBZ (or equivalently over magnetic Bloch momenta). either resolved or combined as appropriate. To visualize gaps uniformly across different $\phi$ we replace discrete levels with a Lorentzian-broadened density of states (DOS),
\begin{gather}
	\rho(E;\phi)=\frac{1}{N_k}\sum_{n,\mathbf{k}}\frac{1}{\pi}\,\frac{\Gamma(\phi)}{\big[E-\varepsilon_{n}(\mathbf{k};\phi)\big]^2+\Gamma(\phi)^2},
\end{gather}
where $N_k$ is the MBZ $k$-mesh size. The broadening $\Gamma(\phi)$ is a plotting parameter chosen to balance visibility and gap resolution; one may take a constant $\Gamma_{\mathrm{vis}}$ for all $\phi$ or an adaptive choice such as $\Gamma(\phi)=\alpha$ times a representative bandwidth at that flux. For a chosen chemical potential (energy cutoff) $\mu$ we define the filling (per magnetic unit cell and per internal flavor, unless otherwise stated) by
\begin{gather}
n(\mu,\phi)=\int_{-\infty}^{\mu}\rho(E;\phi)\,\mathrm{d}E.
\end{gather}
As illustrate in Fig. \ref{fig:wannierplot}, we then plot $n$ versus $\phi=p/q$ and color the points by $\rho(\mu;\phi)$ (or, equivalently, mark local minima of the DOS). Along spectral gaps the DOS is minimal and the graph $n(\phi)$ aligns into nearly straight ``gap lines.’’ These lines obey the Diophantine/Středa relation
\begin{gather}
n(\mu,\phi)=t\,\phi+\nu,\qquad \phi=\frac{p}{q},
\end{gather}
with an integer intercept $\nu$ and an integer slope $t$ that equals the Chern-number difference across the gap; for a gap above a set of fully occupied bands, $t$ equals the total Chern number of those occupied bands. Operationally, one collects the $(\phi,n)$ points where $\rho$ is locally minimal for a given gap and performs a linear fit to extract $(t,\nu)$; repeating this across gaps produces the Landau fan of straight rays with integer slopes. Finally, as a non-visual check, we compute Chern numbers independently from Berry curvature (e.g., by Wilson loops on the same MBZ mesh) and confirm that the slopes $t$ obtained from the fan coincide with the corresponding topological invariants. This procedure yields a gap-by-gap topological labeling from the DOS alone, and, by construction, agrees with the Berry-curvature integral results.

\section{Magnetic hybrid-Wannier method and common-basis projection}

The Landau-level (LL) expansion used in the main text directly resolves the
weak-field splitting.  A complementary hybrid-Wannier (hWF) representation
places the finite-field and zero-field states in one six-orbital plane-wave
basis and thereby enables a momentum-resolved component projection.  The LL
cutoff supplies a systematically controlled spectral reference, while the
variational hWF construction supplies the common-basis wave functions.  Their
spectral agreement validates the target magnetic manifold before projection.

\subsection{Zero-field hWF frame}

For one valley, a zero-field eigenstate of the tMBG continuum Hamiltonian is
expanded as
\begin{equation}
	|\psi_{n\mathbf k}\rangle=
	\sum_{\mathbf G,\lambda}
	c^{(n)}_{\mathbf G\lambda}(\mathbf k)
	|\mathbf k+\mathbf G,\lambda\rangle,
	\label{eq:hw-plane-wave-state}
\end{equation}
where $\lambda$ labels the two sublattices in each of the three graphene
layers.  We retain a multiband active frame $U(\mathbf k)$ and parallel
transport it along $k_1$ using the unitary part of the neighboring-frame
overlap,
\begin{equation}
	L_1(\mathbf k)=\operatorname{polar}\!\left[
	U^\dagger(\mathbf k)U(\mathbf k+\Delta k_1)
	\right].
	\label{eq:hw-link}
\end{equation}
The boundary link includes the reciprocal-lattice sewing matrix.  Its Wilson
loop determines the hWF centers,
\begin{equation}
	\mathcal W_1(k_2)=\prod_{k_1}L_1(k_1,k_2),\qquad
	\mathcal W_1\chi_a=e^{2\pi iX_a(k_2)}\chi_a,
	\label{eq:hw-wilson}
\end{equation}
and Fourier transformation produces states localized along $\mathbf a_1$,
\begin{equation}
	|W_{aR}(k_2)\rangle=
	\frac{1}{\sqrt{N_1}}\sum_{k_1}
	e^{-2\pi ik_1R}|w_a(k_1,k_2)\rangle.
	\label{eq:hw-zero-field-hwf}
\end{equation}
The calculation below retains six zero-field bands, allowing the magnetic
translation orbit to explore the target band and its nearby dispersive
partners while preserving the full continuum description.

\subsection{Magnetic trial states and the complete field term}

At rational flux $\phi/\phi_0=p/q$, magnetic translations satisfy
\begin{equation}
	\hat t_2\hat t_1=e^{2\pi ip/q}\hat t_1\hat t_2.
	\label{eq:hw-magnetic-algebra}
\end{equation}
Following Refs.~\cite{WangVafek2022NarrowBands,WangVafek2023HybridWannier},
we form simultaneous eigenstates of $\hat t_1$ and $\hat t_2^q$ from the
magnetic-translation orbit of a zero-field hWF,
\begin{equation}
	|\Phi_{arn}(\boldsymbol\kappa)\rangle=
	\frac{1}{\sqrt{N_1}}
	\sum_s e^{i\Phi_s(\boldsymbol\kappa,n)}
	\hat t_1^s
	\left|W_{an}\!\left(\kappa_2+\frac rq\right)\right\rangle,
	\label{eq:hw-magnetic-trial}
\end{equation}
where $r=0,\ldots,q-1$ labels the folded momentum sector and $n$ labels
neighboring hWF centers.  The phase contains the magnetic cocycle
$-\pi p_Bs(s-1)/(2q)$, with $p_B=\operatorname{sgn}(B)p$, together with
the Bloch and hWF-center phases.  Equation~
\eqref{eq:hw-magnetic-algebra} fixes this phase consistently across the
magnetic Brillouin zone.

The trial states are nonorthogonal.  Let $V_{\Delta\mathbf G}$ be the
reciprocal-lattice sewing matrix in the six-orbital plane-wave basis.  The
elementary overlap block is
\begin{equation}
	M_{ab}(\mathbf k_r,\mathbf k_c;\Delta\mathbf G)=
	\langle W_a(\mathbf k_r)|V_{\Delta\mathbf G}
	|W_b(\mathbf k_c)\rangle,
	\label{eq:hw-overlap-block}
\end{equation}
and summing these blocks with the phases in
Eq.~\eqref{eq:hw-magnetic-trial} gives the overlap matrix $\mathcal O$.
If $h_W(\mathbf k)$ denotes the zero-field Hamiltonian in the hWF frame,
the corresponding zero-field Hamiltonian block is
\begin{equation}
	H^{(0)}_{ab}(\mathbf k_r,\mathbf k_c)=
	\left[h_W(\mathbf k_r)M(\mathbf k_r,\mathbf k_c)\right]_{ab}.
	\label{eq:hw-h0-block}
\end{equation}
The Hamiltonian acts on the row momentum before the magnetic-scattering
overlap, preserving the ordered row-column construction.

For the Landau gauge used in the magnetic-translation construction,
minimal coupling gives
\begin{equation}
	\hat H(B)=\hat H_0-
	\frac{\operatorname{sgn}(B)}{\ell_B^2}\,x\hat v_y,
	\qquad
	\hat v_y=\frac{\partial H_0}{\partial k_y},
	\label{eq:hw-full-field-hamiltonian}
\end{equation}
where $\ell_B^2=A_Mq/(2\pi p)$.  Since the tMBG continuum Hamiltonian is
linear in momentum, Eq.~\eqref{eq:hw-full-field-hamiltonian} contains the
complete field dependence within this model.
The tMBG velocity contains the extra trigonal-warping and particle-hole
asymmetric couplings of the Bernal bilayer,
\begin{equation}
	v_y=-v_F\,\mathbb 1_3\otimes\sigma_y+
	\begin{pmatrix}
	0&0&0\\
	0&0&D_y^\dagger\\
	0&D_y&0
	\end{pmatrix},\qquad
	D_y=
	\begin{pmatrix}
	iv_3&0\\
	-iv_4&iv_3
	\end{pmatrix}.
	\label{eq:hw-tmbg-velocity}
\end{equation}
Using this full tMBG velocity retains the finite-field matrix elements from
trigonal warping and particle-hole asymmetry.

The $y$ integral fixes the second reciprocal harmonic, while the position
operator couples all first-harmonic pairs.  For a box of length $L_x$, the
required analytic matrix element is
\begin{align}
	I_x(\Delta_x)
	&=\frac1{L_x}\int_{-L_x/2}^{L_x/2}
	xe^{i\Delta_xx}\,dx \nonumber\\
	&=\frac{L_x\Delta_x\cos(\Delta_xL_x/2)
	-2\sin(\Delta_xL_x/2)}
	{iL_x\Delta_x^2},\nonumber\\
	I_x(0)&=0.
	\label{eq:hw-position-integral}
\end{align}
Consequently, the elementary field block has the form
\begin{equation}
	H^{(B)}_{ab}=-\frac{\operatorname{sgn}(B)}{\ell_B^2}
	\sum_{\mathbf G,\mathbf G'}
	I_x(\Delta_x)
	c^{a*}_{\mathbf G}(\mathbf k_r)
	v_yc^b_{\mathbf G'}(\mathbf k_c),
	\label{eq:hw-field-block}
\end{equation}
subject to the second-harmonic selection rule and reciprocal sewing.  All
first-harmonic pairs in Eq.~\eqref{eq:hw-field-block} are retained.

\subsection{Generalized eigenproblem and LL comparison}

Combining the zero-field and field blocks gives
\begin{equation}
	\mathcal H=\mathcal H_0+\mathcal H_B,\qquad
	\mathcal H\mathbf c_{\mu\boldsymbol\kappa}
	=E_{\mu\boldsymbol\kappa}
	\mathcal O\mathbf c_{\mu\boldsymbol\kappa}.
	\label{eq:hw-generalized-eigenproblem}
\end{equation}
We diagonalize $\mathcal O=V\operatorname{diag}(o_j)V^\dagger$ and retain
the physical rank required by the St\v{r}eda count,
\begin{equation}
	N_{\rm phys}=N_{\rm act}q+C_{\rm act}\operatorname{sgn}(B)p.
	\label{eq:hw-physical-rank}
\end{equation}
The orthonormalizing map
$S=V_{\rm keep}\operatorname{diag}(o_j^{-1/2})$ reduces
Eq.~\eqref{eq:hw-generalized-eigenproblem} to the Hermitian problem
$S^\dagger\mathcal HS\,d=Ed$, with $\mathbf c=Sd$.

\begin{figure}
	\centering
	\includegraphics[width=0.82\linewidth]{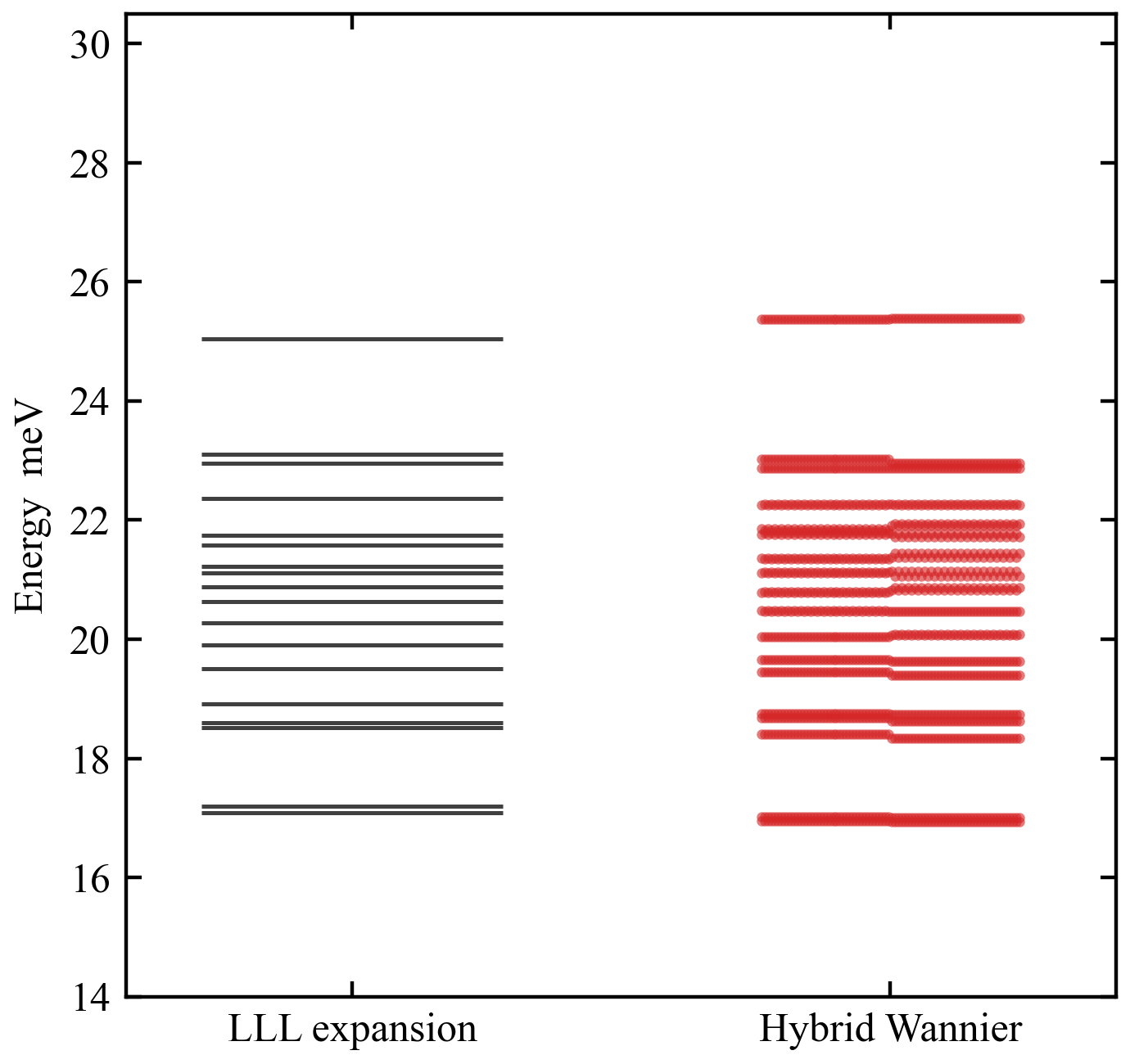}
	\caption{Absolute weak-field energy levels at $p/q=1/20$ and
	$\kappa=0.7$.  Horizontal lines are obtained from the LL expansion,
	while red points show the magnetic hWF spectrum over the magnetic
	Brillouin zone.  The two independent methods reproduce the same target
		manifold and the isolated branch above its internal gap.  Their small
		energy spread quantifies the variational hWF truncation.}
	\label{fig:sm-lll-hw-levels}
\end{figure}

Figure~\ref{fig:sm-lll-hw-levels} compares both spectra on their absolute
energy scale.  The mean levels in the target manifold agree on a scale
well below the separation from remote bands, and both calculations resolve
the same internal gap and upper isolated branch.  We therefore use the hWF
eigenstates for the common-basis projection below, while retaining the LL
calculation as the independent spectral benchmark.

\subsection{Quantum geometry and the trace condition}

The spectrum identifies the isolated magnetic branch, and its momentum-space
wave-function texture determines the corresponding quantum geometry.  We use
the rank-one projector of the isolated upper branch, so that its rotation
relative to the other states in the 18-level target manifold remains part of
the measured geometry.

Because the numerical basis varies with momentum, we first carry the states at
neighboring momenta into a common basis.  Let $U(\mathbf k)$ be an orthonormal
frame of rank $r$, and let $T_{\boldsymbol\delta}$ carry the basis at
$\mathbf k+\boldsymbol\delta$ into the basis at $\mathbf k$.  This gives the
physical, gauge-invariant projector distance
\begin{equation}
	d(\mathbf k,\boldsymbol\delta)=
	r-\left\|U^\dagger(\mathbf k)T_{\boldsymbol\delta}
	U(\mathbf k+\boldsymbol\delta)\right\|_F^2.
	\label{eq:sm-projector-distance}
\end{equation}
For two small physical momentum steps $\boldsymbol\delta_1$ and
$\boldsymbol\delta_2$, let the corresponding distances be $d_1$ and $d_2$,
and let $d_{12}$ be the distance between the two displaced projectors.  To
leading order in the step size,
\begin{equation}
	\begin{aligned}
	D&=
	\begin{pmatrix}
	d_1&(d_1+d_2-d_{12})/2\\
	(d_1+d_2-d_{12})/2&d_2
	\end{pmatrix},\\
	g_{\rm cart}&=B^{-T}DB^{-1}.
	\end{aligned}
	\label{eq:sm-cartesian-metric}
\end{equation}
where $B=(\boldsymbol\delta_1,\boldsymbol\delta_2)$.  This coordinate
transformation incorporates the angle between the two reciprocal directions
of the magnetic mesh.  The trace condition is then evaluated in Cartesian
coordinates as $\operatorname{Tr}g=g_{xx}+g_{yy}$.

The same transported overlaps form a plaquette Wilson loop, whose phase gives
the Berry curvature $\Omega(\mathbf k)$.  Thus the metric and curvature are
calculated from one consistent connection and include the momentum dependence
of both the state and its basis.  The integrated trace excess is
\begin{equation}
	\eta=\int_{\rm MBZ}
	\left[\operatorname{Tr}g(\mathbf k)-|\Omega(\mathbf k)|\right]
	d^2k .
	\label{eq:sm-magnetic-trace-excess}
\end{equation}

For a magnetic hWF calculation, the neighboring-frame overlap naturally
contains a unitary rotation within the retained magnetic space and a positive
amplitude describing the coverage of that space.  Polar decomposition
organizes these two pieces and provides a basis-covariant parallel connection.
At each magnetic momentum, let
\begin{equation}
	\Phi(\mathbf k)=
	\left(|\phi_0(\mathbf k)\rangle,\ldots,
	|\phi_{R-1}(\mathbf k)\rangle\right)
	\label{eq:sm-hwf-correction-frame}
\end{equation}
be an orthonormal frame in the common six-orbital plane-wave basis.  The first
column is the isolated target subband and remains fixed; nearby magnetic
subbands supply the correction space.  For a physical momentum step
$\boldsymbol\delta_\mu$, with reciprocal-boundary sewing included when
necessary, we form
\begin{align}
	\mathcal M_\mu(\mathbf k)
	&=\Phi^\dagger(\mathbf k)T_{\boldsymbol\delta_\mu}
	\Phi(\mathbf k+\boldsymbol\delta_\mu)
	=W_\mu\Sigma_\mu V_\mu^\dagger,\nonumber\\
	\mathcal U_\mu(\mathbf k)&=W_\mu V_\mu^\dagger .
	\label{eq:sm-hwf-polar-connection}
\end{align}
The singular values in $\Sigma_\mu$ monitor the coverage of the finite frame,
whereas the polar factor $\mathcal U_\mu$ gives its unitary parallel
transport.  It is the same overlap orthonormalization that underlies the magnetic hWF
construction~\cite{WangVafek2022NarrowBands,WangVafek2023HybridWannier}.  The
matrix element that transports the target subband is
\begin{equation}
	u_\mu(\mathbf k)=\left[\mathcal U_\mu(\mathbf k)\right]_{00},
	\qquad \widehat u_\mu=\frac{u_\mu}{|u_\mu|}.
	\label{eq:sm-hwf-target-polar-link}
\end{equation}
Under a phase change of the target state, $u_\mu$ acquires the usual endpoint
phases, while unitary rotations among the correction bands preserve it.
Consequently, its normalized phase gives the Berry curvature through the
closed plaquette
\begin{equation}
	\Omega_{\rm hWF}(\mathbf k)=\frac{1}{A_\square}
	\arg\!\left[
	\widehat u_1(\mathbf k)
	\widehat u_2(\mathbf k+\boldsymbol\delta_1)
	\widehat u_1^*(\mathbf k+\boldsymbol\delta_2)
	\widehat u_2^*(\mathbf k)
	\right],
	\label{eq:sm-hwf-intrinsic-curvature}
\end{equation}
where $A_\square=\det(\boldsymbol\delta_1,
\boldsymbol\delta_2)$ is the signed Cartesian plaquette area.

The phase of $u_\mu$ determines the curvature, while its modulus measures how
much the target line rotates into the correction bands.  The corresponding
residual projector distances are
\begin{equation}
	d_\mu^{\rm res}=1-|u_\mu|^2,
	\qquad
	d_{12}^{\rm res}=1-|u_{12}|^2,
	\label{eq:sm-hwf-residual-distances}
\end{equation}
where $u_{12}$ is obtained by the same polar construction between
$\mathbf k+\boldsymbol\delta_1$ and
$\mathbf k+\boldsymbol\delta_2$.  Inserting these distances into
Eq.~\eqref{eq:sm-cartesian-metric} gives the Cartesian residual metric
$g_{ij}^{\rm res}$.  The matrix $B$ in Eq.~\eqref{eq:sm-cartesian-metric}
incorporates the angle between $k_1$ and $k_2$ and directly produces the
physical Cartesian trace.

We next identify the ideal part of the geometry.  For the $C_3$-compatible
complex coordinate $z=x+iy$, an antiholomorphic, or vortexable, rank-one
component saturates the local trace condition.  Its K\"ahler metric is
therefore fixed by the hWF Berry curvature as~\cite{Ledwith2023Vortexability}
\begin{equation}
	g_{ij}^{\rm ideal}(\mathbf k)=
	\frac{|\Omega_{\rm hWF}(\mathbf k)|}{2}\,\delta_{ij},
	\qquad
	\operatorname{Tr}g^{\rm ideal}=|\Omega_{\rm hWF}|.
	\label{eq:sm-hwf-ideal-metric}
\end{equation}
We accordingly decompose the intrinsic hWF geometry as
\begin{align}
	g_{ij}^{\rm int}&=g_{ij}^{\rm ideal}+g_{ij}^{\rm res},\nonumber\\
	\eta_{\rm hWF}^{\rm int}
	&=\int_{\rm MBZ}\operatorname{Tr}g^{\rm res}(\mathbf k)\,d^2k .
	\label{eq:sm-hwf-intrinsic-trace}
\end{align}
Because the first term saturates the trace condition, the integrated trace
excess is the integral of the residual-metric trace in
Eq.~\eqref{eq:sm-hwf-intrinsic-trace}.
A dimensionless measure of the ideal component is
\begin{equation}
	f_{\rm ideal}=
	\frac{\int_{\rm MBZ}|\Omega_{\rm hWF}|\,d^2k}
	{\int_{\rm MBZ}\operatorname{Tr}g^{\rm int}\,d^2k}.
	\label{eq:sm-hwf-ideal-fraction}
\end{equation}
For a projected-density calculation, the same information can be packaged as
a finite-link ideal form factor.  If $\overline g^{\rm ideal}$ denotes the
ideal metric averaged over the two endpoints of a link, we define
\begin{equation}
	F_\mu^{\rm ideal}(\mathbf k)=
	\widehat u_\mu(\mathbf k)
	\exp\!\left[-\frac12
	\delta_\mu^i\overline g_{ij}^{\rm ideal}
	\delta_\mu^j\right].
	\label{eq:sm-hwf-ideal-form-factor}
\end{equation}
Its phase is the hWF Berry connection and its modulus is fixed by the ideal
K\"ahler metric.  It therefore provides the ideal long-wavelength density form
factor as well as its Berry curvature and quantum metric.

We choose a multiband correction frame with $R\geq2$, allowing the target line
to rotate into nearby magnetic subbands.  The controlled result preserves the
target Chern number, keeps the smallest singular value of $\mathcal M_\mu$
finite, and remains stable as both the correction-frame rank and the momentum
mesh are increased.  These conditions define a predictive, parameter-free
calculation from the magnetic hWF states and their common plane-wave embedding.
The resulting decomposition directly provides the quantum-geometric tensor
and projected-density form factors.

The LL calculation below supplies an external finite-field benchmark of the
same rank-one geometry.  In that representation,
$T_{\boldsymbol\delta}$ in Eq.~\eqref{eq:sm-projector-distance} is the analytic
LL form-factor matrix and the Wilson loop also contains the background
Landau-gauge phase.  The hWF construction evaluates the same observables
directly from its polar connection.

For the physical moir\'e flux $p/q=1/20$, the convention of the LL code is
$p_{\rm LL}/q_{\rm LL}=1/10$.  This factor of two follows from the area of
the reciprocal cell used by that implementation and is included before any
comparison is made.  The isolated branch is LL band 82 in the frozen
120-level window and hWF band 79.  With $N_{\rm LL}=400$ and a
$40\times16$ Cartesian LL mesh, we obtain
\begin{equation}
	\begin{aligned}
	C&=-1, & \eta&=0.0988,\\
	\int_{\rm MBZ}\operatorname{Tr}g\,d^2k&=6.38199,
	&\int_{\rm MBZ}|\Omega|\,d^2k&=2\pi.
	\end{aligned}
	\label{eq:sm-ll-geometry-result}
\end{equation}
We orient the two implementations through $C\Omega$, which places
Eq.~\eqref{eq:sm-magnetic-trace-excess} and the displayed curvature in a
common convention.  The normalized fields are nearly uniform:
$\operatorname{std}(C\Omega/l_B^2)=4.43\times10^{-7}$ and
$\operatorname{std}(\operatorname{Tr}g/l_B^2)=3.62\times10^{-7}$.

\begin{figure*}[t]
	\centering
	\includegraphics[width=0.95\textwidth]{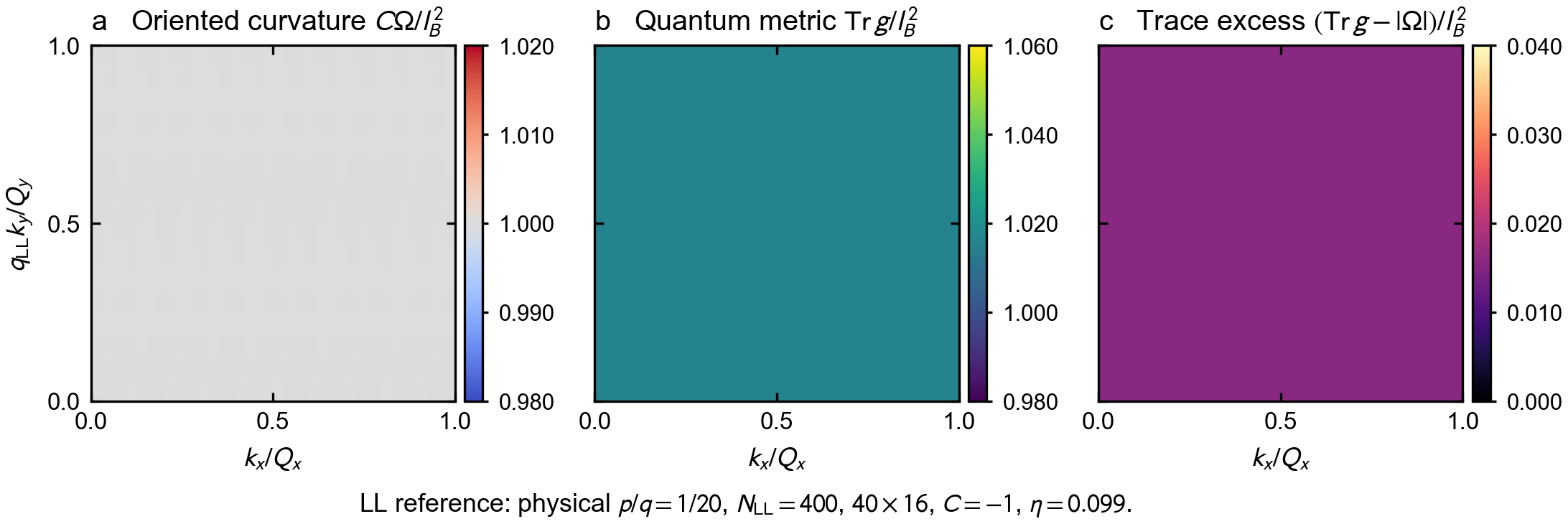}
	\caption{Quantum geometry of the isolated upper magnetic subband from the
	converged LL expansion at physical flux $p/q=1/20$ and $\kappa=0.70$.
	(a) Berry curvature oriented as $C\Omega/l_B^2$, so that the displayed
	mean is positive independent of the momentum-orientation convention.
	(b) Cartesian trace of the quantum metric.  (c) Local trace excess.  The
		fixed physical color scales display the residual discretization at its
		actual amplitude.  The integrated value is $\eta=0.0988$.}
	\label{fig:sm-ll-quantum-geometry}
\end{figure*}

The LL cutoff and momentum mesh are converged together.  At
$N_{\rm LL}=400$, refinement from $20\times8$ to $40\times16$ gives
$\eta=0.0331$ and $0.0988$, while a second-order continuum extrapolation gives
$\eta=0.1164$.  Equations~\eqref{eq:sm-hwf-polar-connection}--
\eqref{eq:sm-hwf-intrinsic-trace} provide the intrinsic hWF geometry, and
Eq.~\eqref{eq:sm-ll-geometry-result} supplies the external finite-field
benchmark.  The common-basis construction below then uses the independently
validated hWF eigenstates for the component projection.

\subsection{Common plane-wave basis and gauge transformation}

Let $\mathcal I$ denote the complete magnetic Bloch subspace above the
internal gap and below the next dispersive magnetic band.  Its projector is
\begin{equation}
	P_{\mathcal I}^{B}=
	\sum_{\boldsymbol\kappa}
	\sum_{\mu\in\mathcal I}
	|\Psi^B_{\mu\boldsymbol\kappa}\rangle
	\langle\Psi^B_{\mu\boldsymbol\kappa}|.
	\label{eq:hw-isolated-projector}
\end{equation}
The spectral gap defines the selected magnetic subspace before its
momentum-resolved components are evaluated.

To unfold Eq.~\eqref{eq:hw-isolated-projector}, we first combine the
orthonormalizing map and magnetic eigenvectors,
$C_{\mathcal{I}}=SD_{\mathcal{I}}$.  We then coherently sum all hWF centers,
folded sectors, magnetic
translations, and reciprocal sewing matrices.  The resulting common-basis
amplitude can be written schematically as
\begin{equation}
	A_{\mathbf G\lambda,\mu}(\mathbf k)=
	\sum_{t,r,n,a}e^{i\varphi_{trn}}
	\left[V_{\Delta\mathbf G}W(\mathbf k_{tr})\right]_{\mathbf G\lambda,a}
	[C_{\mathcal{I}}]_{arn,\mu}.
	\label{eq:hw-common-basis-lift}
\end{equation}
The coherent sum retains all complex phases across the $q$ folded momenta and
therefore implements the magnetic-translation representation.  A final
Lowdin correction using the Gram matrix
\begin{equation}
	G_{\mu\nu}=\sum_{\mathbf k,\mathbf G,\lambda}
	A^*_{\mathbf G\lambda,\mu}(\mathbf k)
	A_{\mathbf G\lambda,\nu}(\mathbf k)
	\label{eq:hw-common-gram}
\end{equation}
restores orthonormality of the selected magnetic subspace in the truncated
common plane-wave space while preserving the subspace itself.

The trial states in Eq.~\eqref{eq:hw-magnetic-trial} are constructed in
Landau gauge, and the zero-field comparison uses a rotation-covariant
symmetric gauge.  We connect them by the single unitary
transformation
\begin{equation}
	U_{L\rightarrow S}(\mathbf r)=
	\exp\!\left[-i\operatorname{sgn}(B)
	\frac{xy}{2\ell_B^2}\right]
	\label{eq:hw-gauge-transform}
\end{equation}
to Eq.~\eqref{eq:hw-common-basis-lift}.  Numerically this is evaluated by a
Fourier transform to the common real-space grid, multiplication by
Eq.~\eqref{eq:hw-gauge-transform}, and transformation back.  The displayed
momentum texture retains the result of this single gauge transformation in its
unsymmetrized form.

Finally, we recover separately labeled energy bands from the smooth
non-Abelian frame used to construct the hWFs.  If $F(\mathbf k)$ is that frame,
we rediagonalize its Bloch Hamiltonian,
\begin{equation}
	F^\dagger H_0F\,V_E=V_E\varepsilon,\qquad
	U_E(\mathbf k)=F(\mathbf k)V_E(\mathbf k),
	\label{eq:hw-energy-frame-recovery}
\end{equation}
and project the transformed magnetic amplitudes onto the zero-field energy
eigenstates $U_E$.  This produces continuous physical energy-band labels from
the smooth frame.

\subsection{Target-band weight}

For a zero-field band $n$, the momentum-resolved component is
\begin{equation}
	W_n(\mathbf k)=
	\frac{1}{N_{\rm MBZ}N_{\mathcal I}}
	\sum_{\boldsymbol\kappa,\mu\in\mathcal I}
	\left|
	\langle u^0_{n\mathbf k}|
	U_{L\rightarrow S}|\Psi^B_{\mu\boldsymbol\kappa}\rangle
	\right|^2.
	\label{eq:hw-target-component}
\end{equation}
The prefactor normalizes the trace of the discrete spectral projector and sets
the probability scale at each momentum-grid point while preserving the
texture.  For visualization in
Fig.~\ref{main-fig:weakfieldcomponenthex}, we therefore plot
\begin{equation}
	\overline W_{\rm target}(\mathbf k)=
	\frac{W_{\rm target}(\mathbf k)}
	{\max_{\mathbf k}W_{\rm target}(\mathbf k)},
	\label{eq:hw-relative-component}
\end{equation}
using the natural minimum and maximum of the data.

Within the central low-energy pair, $91.9\%$ of the projected weight returns
to the original target band.  The isolated magnetic branch is therefore
predominantly inherited from the zero-field target band.  Its momentum
distribution is nevertheless strongly nonuniform: all three equivalent
selected-valley $K$ regions carry much less weight than the smoother parts of
the Brillouin zone.  The unsymmetrized distribution is consistent with the
concentration of rapid Bloch-state variation and
quantum metric near $K$ in Fig.~\ref{main-fig:bandstructure}.  This comparison
provides a qualitative geometric interpretation.  Figure~
\ref{fig:sm-ll-quantum-geometry} supplies the quantitative finite-field LL
benchmark, and Eqs.~\eqref{eq:sm-hwf-polar-connection}--
\eqref{eq:sm-hwf-intrinsic-trace} define the intrinsic hWF diagnostic.  The
component projection uses the common-basis spectral projector constructed in
Eqs.~\eqref{eq:hw-isolated-projector}--\eqref{eq:hw-target-component}.
The implementation of the weak-field ideal-component analysis described in
this section is publicly available in Ref.~\cite{SongChang2026MoireFCICode}.

\bibliography{ColorFCI.bib}